\documentclass[a4paper,fleqn]{cas-dc}

\usepackage{enumitem}
\usepackage[numbers]{natbib}
\usepackage[figuresright]{rotating}
\usepackage{moreverb}
\usepackage{amsmath,amssymb,amsfonts}
\usepackage{algorithmic}
\usepackage[inkscapelatex=false]{svg}
\usepackage{url}
\usepackage{subfig}
\usepackage{subcaption}
\usepackage{hyperref}
\hypersetup{
    colorlinks=true,
    linkcolor=blue,
    citecolor=blue,
    urlcolor=blue,
    filecolor=blue,
    linktoc=all,
}
\usepackage{tabularx}
\usepackage{pifont}
\usepackage{fontawesome}
\usepackage{longtable}
\usepackage{graphicx}
\usepackage{textcomp}
\usepackage{xcolor}
\usepackage{tcolorbox}
\usepackage{ragged2e}
\usepackage{balance}
\usepackage{colortbl}
\usepackage{multirow}
\usepackage{booktabs}
\usepackage{CJKutf8}
\usepackage{siunitx}
\usepackage{rotfloat}
\usepackage{float}
\usepackage{lscape}
\usepackage{mathrsfs}
\usepackage{adjustbox}
\usepackage{afterpage}
\usepackage{caption}
\usepackage{float}
\usepackage[section]{placeins}

\usepackage{placeins}

\usepackage{xfp}   

\def\tsc#1{\csdef{#1}{\textsc{\lowercase{#1}}\xspace}}
\tsc{WGM} \tsc{QE} \tsc{EP} \tsc{PMS} \tsc{BEC} \tsc{DE}

\definecolor{lightgreen}{RGB}{255,255,255} 

\begin{document}
\begin{sloppypar}

\let\WriteBookmarks\relax
\def\floatpagepagefraction{1}
\def\textpagefraction{.001}

\shorttitle{Assessing Small Language Models for Code Generation}
\shortauthors{MM Hasan et al.}

\title [mode = title]{Assessing Small Language Models for Code Generation: An Empirical Study with Benchmarks}

\author[tuni]{Md Mahade Hasan}
\ead{mdmahade.hasan@tuni.fi}

\author[tuni]{Muhammad Waseem}
\ead{muhammad.waseem@tuni.fi}

\author[tuni]{Kai-Kristian Kemell}
\ead{kai-kristian.kemell@tuni.fi}

\author[tuni]{Jussi Rasku}
\ead{jussi.rasku@tuni.fi}

\author[tuni]{Juha Ala-Rantala}
\ead{juha.ala-rantala@tuni.fi}

\author[tuni]{Pekka Abrahamsson}
\ead{pekka.abrahamsson@tuni.fi}

\address[tuni]{Faculty of Information Technology and Communication Sciences, Tampere University, Tampere, Finland}

\begin{abstract}
The recent advancements of Small Language Models (SLMs) have opened new possibilities for efficient code generation. SLMs offer lightweight and cost-effective alternatives to Large Language Models (LLMs), making them attractive for use in resource-constrained environments. However, empirical understanding of SLMs, particularly their capabilities, limitations, and performance trade-offs in code generation remains limited. This study presents a comprehensive empirical evaluation of 20 open-source SLMs ranging from 0.4B to 10B parameters on five diverse code-related benchmarks (HumanEval, MBPP, Mercury, HumanEvalPack, and CodeXGLUE). The models are assessed along three dimensions: i) functional correctness of generated code, ii) computational efficiency and iii)  performance across multiple programming languages. The findings of this study reveal that several compact SLMs achieve competitive results while maintaining a balance between performance and efficiency, making them viable for deployment in resource-constrained environments. However, achieving further improvements in accuracy requires switching to larger models. These models generally outperform their smaller counterparts, but they require much more computational power. We observe that for 10\% performance improvements, models can require nearly a 4x increase in VRAM consumption, highlighting a trade-off between effectiveness and scalability. Besides, the multilingual performance analysis reveals that SLMs tend to perform better in languages such as Python, Java, and PHP, while exhibiting relatively weaker performance in Go, C++, and Ruby. However, statistical analysis suggests these differences are not significant, indicating a generalizability of SLMs across programming languages. Based on the findings, this work provides insights into the design and selection of SLMs for real-world code generation tasks. 

\end{abstract}


\maketitle

\section{Introduction}
\label{Introduction}

Small Language Models (SLMs) are lightweight variants of Large Language Models (LLMs), designed for efficiency, and ease of use in resource-constrained environments \cite{xu2024survey}. Unlike LLMs, which contain hundreds of billions or even trillions of parameters, SLMs typically range from 1 million to 10 billion parameters \cite{Wang2024ACS}. Despite being significantly smaller, SLMs have shown effective performance on domain-specific tasks such as code generation \cite{Subramanian2025survey}. This has led to growing interest in SLMs, as they offer a good balance between performance and efficiency \cite{Nguyen2024survey}. With lower memory usage, faster inference, and easier deployment, SLMs can be adopted as a practical and sustainable alternative \cite{schick-schutze-2021-just}. Therefore, to support SLMs further development and successful adoption in the real world, it is necessary to assess their performance. 

Although SLMs are getting more attention, our understanding of their performance across code generation remains limited. Most studies so far have focused on LLMs such as Codex \cite{Chen2021EvaluatingLL} and CodeLlama \cite{CodeLlama}, while leaving smaller models relatively underexplored. Moreover, the few existing results often lack consistency in terms of benchmarks used (e.g., HumanEval, MBPP), programming language diversity (e.g., Python, Java, C++), and reporting of computational efficiency (e.g., memory usage, inference speed). This makes it difficult to compare results or draw meaningful conclusions about their effectiveness. Without clear testing, it is challenging to develop a complete and reliable understanding of SLMs and their real-world capabilities.

Despite recent progress in the development of SLMs, the lack of comprehensive evaluations leaves many open questions about their actual effectiveness in practice \cite{Lu2024SmallLM}. Existing studies mostly focus on general natural-language processing tasks \cite{zellers2019hellaswag} \cite{hendrycks2020measuring} \cite{lewkowycz2022solving}. As a result, the effectiveness of SLMs for specialised software engineering tasks, such as code generation, remains insufficiently understood. Besides, the absence of cross-programming-language evaluations also limits understanding of whether SLM performance and efficiency vary across different programming languages. Moreover, prior work does not systematically examine task-specific trade-offs between code-generation quality and computational efficiency, including latency and memory usage, which are critical for practical deployment \cite{Lepagnol2024SmallLM}. These gap motivates our study. To address these, we conduct a systematic evaluation aimed at building a better understanding of how these models perform across a range of conditions for practical usage. Specifically, we focus on three key aspects: (i) performance of SLMs on code generation, (ii) how they balance computational efficiency and performance, and (iii) how they perform in different programming languages. By examining these dimensions, we aim to provide insights into the strengths, weaknesses, and practical trade-offs of SLMs. 

Considering the research gap and our research objectives, we have formulated the following three questions along with their rationale:

\begin{itemize}
 \label{research_question}
 \item \textbf{RQ1}: How do different small language models perform on code generation tasks across established benchmarks?

 \textbf{Rationale}: This question evaluates the effectiveness of small language models in code generation by analysing their performance across various benchmarks. The goal is to identify their strengths and limitations to guide future model development and practical applications.
 
 \item \textbf{RQ2}: What are the trade-offs between computational efficiency and practical usability of small language models for code generation?

 \textbf{Rationale}: This question examines how small language models balance performance with resource usage, such as memory and inference speed. Understanding these trade-offs helps determine which models are best suited for real-world scenarios with hardware or time constraints.
 
 \item \textbf{RQ3}: Do Small Language Models perform consistently across different programming languages, or are they better suited to some languages than others?

 \textbf{Rationale}: This question explores whether small language models perform consistently across different programming languages or favour some over others. Identifying language-specific patterns can guide improvements in training and enhance their reliability for multilingual coding tasks.

\end{itemize}

To address these research questions, we designed and conducted an empirical benchmark-based evaluation of multiple open-source SLMs. Our key contributions are:

\begin{itemize}
 \item A systematic evaluation of 20 open-source SLMs (i.e., 0.4B–10B parameters) divided into three groups across five widely used benchmarks, HumanEval, MBPP, Mercury, HumanEvalPack, and CodeXGLUE—using a unified decoding and prompting setup.
 
 \item Detailed performance measurements of selected SLMs, including functional correctness \textit{(pass@k)}, GPU memory consumption, and inference speed, to better understand the computational trade-offs inherent in these models.

 \item Evaluation of language-specific performance across selected SLMs, assessing the multilingual capabilities and limitations of small language models in code generation tasks.

\end{itemize}

Our results show that larger SLMs from Group 3 (parameter size $>$3B to $\leq$10B) generally deliver the highest accuracy, but several smaller and mid-sized models from Groups 1 and 2 (parameter size $\leq$3B) perform competitively while requiring substantially fewer resources. VRAM usage increases sharply with model size, although inference times remain statistically similar. While multilingual performance varies across individual languages, these differences are not statistically significant within the language set included in our benchmarks. Overall, the findings indicate important trade-offs between accuracy and efficiency that practitioners must consider when selecting SLMs for practical deployment.

\textbf{Paper Organization}: Section~\ref{Methodology} outlines the research methodology, which describes the experimental design, including selected models, benchmarks, and evaluation metrics. Section~\ref{Experimental Results} presents the experimental findings, while Section~\ref{Discussion} discusses their implications. Section~\ref{ThreatstoValidity} addresses potential threats to the validity of the results of this study. Section~\ref{RelatedWork} reviews related work. Section~\ref{Conclusion} summarizes the contributions and concludes the paper.



\section{Methodology}
\label{Methodology}

\subsection{Research Design Overview}

\begin{figure*}[!htp]
    \centering
    \includegraphics[width=\linewidth]{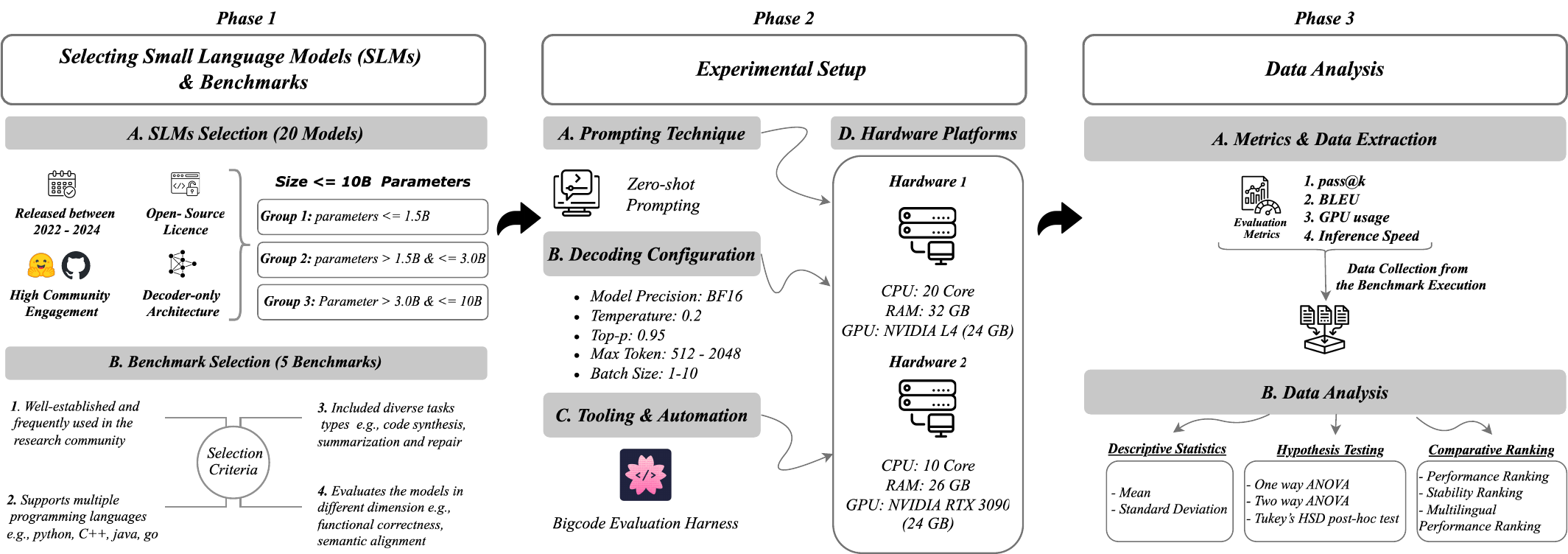}
    \caption{Overview of our research design and evaluation workflow.}
    \label{fig:research_design}
\end{figure*}

This study adopts an empirical, benchmark-based methodology to evaluate SLM performance in code generation. The workflow consists of three phases: (i) selecting SLMs \& Benchmarks, (ii) Experimental Setup, and (iii) Data Analysis. This design directly supports our three research questions on model performance, computational efficiency, and multilingual capability, and ensures reproducible comparisons across five benchmarks. Figure~\ref{fig:research_design} provides an overview of this process. Phase 1 covers the selection of 20 open-source SLMs and five benchmarks based on release date, licence, architecture, community support, and language coverage. Phase 2 illustrates the unified experimental setup, including zero-shot prompting, decoding configurations, hardware platforms, and automated evaluation. Phase 3 summarises data extraction and analysis, including metric computation (pass@k, BLEU, VRAM, inference time) and statistical testing. Together, these phases form a consistent and structured evaluation pipeline.

\subsection{SLMs Selection }
We systematically selected 20 SLMs for this study using the following criteria. i) We selected SLMs released between 2022 and 2024 to ensure the study captures the latest advancements in the field. ii) We prioritised SLMs with high community engagement, using metrics such as Huggingface downloads and GitHub stars to reflect popularity. iii) We included only SLMs with open-source licenses (e.g., MIT, Apache 2.0) to support accessibility. iv) We restricted our selection to SLMs with up to 10 billion parameters\cite{Wang2024ACS} and categorised them into three groups, i.e., a) SLMs parameter size $\leq$1.5B, b) SLMs parameter size $>$1.5B to $\leq$ 3B, and c) SLMs parameter size $>$3B to $\leq$10B. v) We considered only SLMs that follow a decoder-only transformer architecture, which is widely used in autoregressive code generation tasks due to its effectiveness and simplicity \cite{svyatkovskiy2020intellicode}. Importantly, all our selected models are specifically designed for code generation tasks, ensuring relevance to the study’s focus on code generation performance. Table~\ref{tab:slm-overview} provides an overview of the selected SLMs, including their specifications and metadata.

\renewcommand{\arraystretch}{1.5} 

\begin{table*}[!htp]
\centering
\caption{Overview of Selected Small Language Models (SLMs)}
\label{tab:slm-overview}
\resizebox{\textwidth}{!}{%
\begin{tabular}{|l|c|c|c|l|c|c|}
\hline
\rowcolor{lightgray}
\textbf{Model\_Parameter} & 
\textbf{Architecture} & 
\textbf{License} & 
\textbf{Release Date} & 
\textbf{Company} & 
\shortstack{\textbf{Huggingface Downloads} \\ \textbf{(Feb 2025)}} &
\shortstack{\textbf{GitHub Stars} \\ \textbf{(Feb 2025)}} \\
\hline

\multicolumn{7}{|c|}{\rule{0pt}{2.6ex}\textbf{Group 1: SLMs parameter size $\leq$1.5B}\rule[-1.2ex]{0pt}{0pt}} \\
\hline

PolyCoder\_0.4B &
Decoder & MIT & Feb 2022 & Carnegie Mellon University & 273 & 1.8K \\
\hline

InCoder\_1.3B &
Decoder & CC-BY-NC-4.0 & Apr 2022 & Facebook/Meta & 3060 & 3.2K \\
\hline

Phi1\_1.3B &
Decoder & MIT & Jun 2023 & Microsoft & 7449 & N/A \\
\hline

DeepSeek-Coder\_1.3B &
Decoder & DEEPSEEK-license & Nov 2023 & DeepSeek-AI & 123488 & 21K \\
\hline

OpenCodeInterpreter\_1.3B &
Decoder & Apache 2.0 & Feb 2024 & MAP & 555 & 1.6K \\
\hline

Yi-Coder\_1.5B &
Decoder & Apache 2.0 & Sep 2024 & 01.AI & 8913 & 7.8K \\
\hline

Qwen2.5-Coder\_1.5B &
Decoder & Apache 2.0 & Sep 2024 & Alibaba Cloud & 11799 & 4.6K \\
\hline

\multicolumn{7}{|c|}{\rule{0pt}{2.6ex}\textbf{Group 2: SLMs parameter size $>$1.5B to $\leq$3B}\rule[-1.2ex]{0pt}{0pt}} \\
\hline

CodeGemma\_2.0B &
Decoder & Gemma & Apr 2024 & Google & 37008 & N/A \\
\hline

PolyCoder\_2.7B &
Decoder & MIT & Feb 2022 & Carnegie Mellon University & 330 & 1.8K \\
\hline

Stable Code\_3.0B &
Decoder & Apache 2.0 & Jan 2024 & Stability AI & 175 & 125 \\
\hline

StarCoder2\_3.0B &
Decoder & BigCode OpenRAIL-M & Feb 2024 & BigCode & 1468288 & 1.9K \\
\hline

Qwen2.5-Coder\_3.0B &
Decoder & Qwen Research & Sep 2024 & Alibaba Cloud & 13540 & 4.6K \\
\hline

\multicolumn{7}{|c|}{\rule{0pt}{2.6ex}\textbf{Group 3: SLMs parameter size $>$3B to $\leq$10B}\rule[-1.2ex]{0pt}{0pt}} \\
\hline

InCoder\_6.7B &
Decoder & CC-BY-NC-4.0 & Apr 2022 & Facebook/Meta & 640 & 3.0K \\
\hline

DeepSeek-Coder\_6.7B &
Decoder & DEEPSEEK-license & Nov 2023 & DeepSeek-AI & 28065 & 21K \\
\hline

OpenCodeInterpreter\_6.7B &
Decoder & Apache 2.0 & Feb 2024 & MAP & 136 & 1.5K \\
\hline

CodeLlama\_7.0B &
Decoder & Llama2 & May 2024 & Meta & 180516 & 16.2K \\
\hline

StarCoder2\_7.0B &
Decoder & BigCode OpenRAIL-M & Feb 2024 & BigCode & 125386 & 2.0K \\
\hline

CodeGemma\_7.0B &
Decoder & Gemma & Apr 2024 & Google & 42304 & N/A \\
\hline

Qwen2.5-Coder\_7.0B &
Decoder & Apache 2.0 & Sep 2024 & Alibaba Cloud & 64152 & 4.6K \\
\hline

Yi-Coder\_9.0B &
Decoder & Apache 2.0 & Sep 2024 & 01.AI & 49711 & 7.8K \\
\hline

\end{tabular}%
}
\end{table*}

\subsection{Benchmark Selection}

\renewcommand{\arraystretch}{1.3} 

\begin{table*}[!t]
\centering
\caption{Overview of the selected benchmarks}
\label{tab:benchmark-overview}
\begin{tabular}{|l|l|l|l|l|}
\hline
\rowcolor{lightgray}
\textbf{Benchmark} & \textbf{Tasks} & \textbf{Language(s)} & \textbf{Metric} & \textbf{Linked RQ(s)} \\ \hline

HumanEval & Code Synthesis & Python & pass@k & RQ1 \\ \hline
MBPP & Basic Programming Problems & Python & pass@k & RQ1 \\ \hline
Mercury & Code Synthesis + Efficiency & Python & pass@k & RQ1 \\ \hline
CodeXGLUE (C2T) & Code-to-Text Summarization & Python, Javascript, Java, PHP, Ruby, Go & BLEU & RQ3 \\ \hline
HumanEvalPack & Synthesis, Repair & Python, Javascript, Java, C++ & pass@k &  RQ3 \\ \hline

\end{tabular}
\end{table*}

We systematically selected five benchmarks, based on the following considerations: i) We prioritised benchmarks that are well-established in the research community and frequently used for evaluating LMs on code-related tasks \cite{10606356, he-etal-2024-ultraeval}. This ensures that our results are comparable with previous studies (e.g., Li \textit{et al.} \cite{li2023starcoder}, Zhuo \textit{et al.} \cite{zhuo2024bigcodebench}) and can be trusted by the broader research community. ii) We ensured diversity in evaluation tasks, including code synthesis, code summarisation, and code repair, to have a comprehensive assessment of SLMs capabilities. This variety reflects the multifaceted nature of software development, enabling us to assess how the SLMs handle different types of code generation challenges. iii) We selected benchmarks that support multiple programming languages, including Python, Java, JavaScript, PHP, Ruby, Go, and C++, to examine multilingual performance. iv) We included benchmarks that assess the SLMs outputs across various dimensions, such as functional correctness, and alignment with reference answers. These dimensions provide a holistic understanding of SLMs ability, balancing both performance and practical utility in real-world coding scenarios. Table~\ref{tab:benchmark-overview} presents an overview of the five selected benchmarks, highlighting their key characteristics and relevance to the research questions.

\subsection{Experimental Setup} \label{sec:experimental_setup}

A controlled experimental setup was established to ensure consistent, fair, and reproducible evaluation of the selected SLMs across the benchmarks. This setup encompassed various key components, including prompting techniques, decoding configurations, hardware platforms, and automation tools, all of which were designed to ensure a comprehensive and unbiased evaluation process.

\subsubsection{Prompting Technique}
We adopted a zero-shot prompting strategy for all benchmarks, where each model received only the raw benchmark input without supplementary examples or contextual hints. Zero-shot prompting is the standard evaluation protocol in widely used code generation benchmarks such as HumanEval \cite{Chen2021EvaluatingLL} and MBPP \cite{austin2021program}, and is also commonly used in recent evaluations of code models (e.g., \cite{li2023starcoder}). Before finalizing this approach, we considered whether alternative prompting strategies (e.g., few-shot prompting) would be appropriate for this study. However, prior work suggests that few-shot or example-driven prompting is primarily enabled by instruction tuning \cite{ouyang2022training}, and that base models do not reliably benefit from demonstration-based prompts. All SLMs evaluated in this study are base models, so incorporating few-shot prompts would introduce methodological bias and reduce comparability across models. For this reason, we intentionally excluded few-shot prompting from the experimental design. We acknowledge, however, that this choice limits our ability to assess model behavior in interactive or example-driven workflows, a limitation further discussed in \autoref{ThreatstoValidity}.

\subsubsection{Decoding Configuration}
Consistent decoding parameters were maintained across all tasks (see Table~\ref{tab:decoding_params}). Adjustments were made only when certain benchmarks needed longer context lengths. To promote deterministic yet slightly different outputs, the temperature was set to 0.2, top-k sampling was disabled by setting it to 0, and nucleus sampling \cite{holtzman2019curious} was applied with top-p set to 0.95. The maximum generation length was fixed at 512 tokens for most tasks; however, for benchmarks demanding longer contexts, such as Mercury and HumanEvalPack (fixtests, fixdocs), this limit was extended to 2048 tokens. A default batch size of 10 was chosen to utilise all the available hardware resources. However, it was reduced to 5 for memory-intensive tasks to prevent out-of-memory errors. For CodeXGLUE, a batch size of 1 was adopted following its official evaluation guidelines. When computing \textit{pass@k}, for all benchmarks we generated 10 samples\textit{(n = 10)} samples per task. Multiple samples are required because SLMs outputs are nondeterministic, and using a single sample would produce unstable estimates of functional correctness. The choice of n = 10 follows the standard configuration used in the official evaluation guidelines, ensuring comparability with prior work while keeping the computational cost manageable. To support full reproducibility, all scripts used for running experiments and analysis are publicly available at: \textit{https://github.com/GPT-Laboratory/Assessing-Small-Language-Models-SLMs-for-Code-Generation.}

\renewcommand{\arraystretch}{1.3} 

\begin{table}[!htbp]
\centering
\caption{Summary of decoding parameters used across all benchmarks.}
\begin{tabular}{|p{0.3\linewidth}|p{0.62\linewidth}|}
\hline
\textbf{Parameter} & \textbf{Value / Description} \\
\hline
Temperature & 0.2 \\
\hline
Top-k & 0 (disabled) \\
\hline
Top-p & 0.95 \\
\hline
Max generation length & 
\begin{tabular}[t]{@{}l@{}}
$\bullet$ 512 tokens (default) \\
$\bullet$ 2048 tokens for Mercury and \\ HumanEvalPack fixtests/fixdocs
\end{tabular} \\
\hline
Batch size & 
\begin{tabular}[t]{@{}l@{}}
$\bullet$ 10 (default across most tasks) \\
$\bullet$ 5 for memory-intensive tasks \\
$\bullet$ 1 for CodeXGLUE (official guidelines)
\end{tabular} \\
\hline
Number of samples (n) for pass@k & 10 \\
\hline

\hline
\end{tabular}
\label{tab:decoding_params}
\end{table}

\subsubsection{Hardware Platforms}
Experiments were conducted in two distinct hardware environments to assess computational efficiency under different resource conditions and reduce total inference time. The first setup featured an Intel Core i7-14700 CPU with 20 cores, 32 GB RAM, and a workstation-grade NVIDIA L4 GPU with 24 GB VRAM. The second setup included an Intel Core i9-10900X CPU with 10 cores, 26 GB RAM, and a consumer-grade NVIDIA RTX 3090 GPU, also featuring 24 GB of VRAM. Models were evenly split between the two environments to balance the computational load and accelerate the evaluation process. Although no direct performance comparison was made between the setups, using both allowed the evaluation to explore different hardware scenarios, thereby offering a deeper understanding of computational resource requirements.

\subsubsection{Tooling and Automation}
We used the BigCode Evaluation Harness\footnote{\url{https://github.com/bigcode-project/bigcode-evaluation-harness}}, an open-source benchmarking framework built on top of Hugging Face Transformers and Accelerate to ensure a consistent, scalable, and reproducible evaluation workflow across all 20 SLMs. The framework provides a unified interface for running code-generation benchmarks and automates several parts of the evaluation pipeline. In our setup, it was responsible for loading models and tokenizers, formatting prompts according to each benchmark’s specification, executing model inference, and computing metrics such as \textit{pass@k} and BLEU score \cite{papineni2002bleu}. By standardizing these steps, the framework ensures that all models are evaluated under identical conditions.

For acceleration, the framework relies on the HuggingFace Accelerate backend, which manages device placement, memory allocation, and mixed-precision execution. All models in this study were executed on a single GPU using PyTorch with BF16 precision, which reduces memory usage while maintaining numerical stability during generation. Accelerate automatically handles casting, GPU memory safety checks, and synchronised execution without requiring manual optimisation. Importantly, we did not employ additional performance-enhancing frameworks such as vLLM, TensorRT, FlashAttention, quantisation, or custom fused kernels. This ensures that all models were evaluated using the same baseline acceleration settings, avoiding discrepancies that could arise from specialised optimisation techniques.


\subsection{Metrics and Data Collection}
In order to answer the three research questions of this study, we used a combination of standard benchmark metrics and system-level performance measurements. Each metric was chosen to align with the objectives of the corresponding research question. 

For RQ1, which investigates the code generation performance of small language models across different benchmarks, we relied on functional correctness metrics. Specifically, we used the \textbf{\textit{pass@k}} metric on HumanEval, MBPP, and Mercury. This metric measures the proportion of problems for which at least one of the top-$k$ generated code completions passes all the reference test cases. It is formally defined as:

\begin{equation}
\text{pass@k} = 1 - \prod_{i=1}^{k} \left(1 - \frac{n_{\text{correct}}}{n} \right)
\end{equation}

where:
\begin{itemize}
    \item $n$ is the total number of generated samples per task,
    \item $n_{\text{correct}}$ is the number of correct samples,
    \item $k$ is the number of samples considered in the top-$k$ evaluation.
\end{itemize}

This metric approximates the probability that at least one of the top-$k$ completions is correct, given a total of $n$ samples. It is particularly useful in evaluating stochastic code generation, where multiple completions are sampled.

To answer RQ2, which focuses on computational efficiency and practical usability, we collected system-level performance metrics during inference. This includes: 

\begin{itemize}
    \item Peak memory usage, measured in gigabytes, to assess resource demands.
     \item Latency, or average time per generation, which captures how quickly a model responds to a single task.
\end{itemize}

These metrics were recorded during benchmarking runs for each model on both hardware configurations to capture variations across computational environments. 

For RQ3, which explores language-specific model performance, we analysed benchmark results by programming language. HumanEvalPack and CodeXGLUE contain tasks written in multiple languages. In HumanEvalPack, models are instructed to generate code, on the other hand, for CodeXGLUE, models are generating code documentation in natural language. For each tasks in each programming language in HumanEvalPack, we computed the same \textit{pass@k} and in CodeXGLUE, we calculated the BLEU score. BLEU is a precision-based metric commonly used in natural language generation to measure the overlap between generated and reference sequences. In the multilingual evaluation, the big-code evaluation harness compute the BLEU score following the standard n-gram precision–based formulation with a brevity penalty. For a candidate generation \(C\) and reference solution \(R\), the BLEU score is calculated as:

\begin{equation}
\text{BLEU} = \text{BP} \cdot \exp\left( \sum_{n=1}^{4} w_n \log p_n \right)
\end{equation}

\begin{itemize}
    \item  where \(p_n\) denotes the modified n-gram precision, \(w_n = \frac{1}{4}\) is the uniform weight for each n-gram order, and BP is the brevity penalty:
    
    \[
        \text{BP} =
        \begin{cases}
        1 & \text{if } |C| > |R|\\
        \exp\left(1 - \frac{|R|}{|C|}\right) & \text{if } |C| \le |R|
        \end{cases}
    \]
    
\end{itemize}

In our context, it quantifies how closely the generated code documentation matches the human-written descriptions, making it suitable for evaluating code-to-text tasks across different languages. This allowed us to examine whether models perform consistently across languages or exhibit biases.

\subsection{Data Analysis}
\label{sub:data_analysis}


The analysis of results in this study involved a combination of descriptive statistical techniques (e.g., mean, median, and standard deviation), hypothesis testing, and comparative ranking \cite{inbook}. To effectively interpret the experimental outcomes, we utilised a range of visualisations, including line graphs, box plots, and bubble plots, to illustrate the various results obtained.

To answer RQ1, we evaluated the code generation performance of SLMs using three standard code generation benchmarks: HumanEval, MBPP, and Mercury. Our evaluation focuses on two key performance aspects: (i) the functional correctness of the generated code, and (ii) the variation in functional correctness across the benchmarks. Functional correctness was measured using the \textit{pass@k} metric for k=1,2,5,10. To summarise each model’s overall performance, we aggregated the \textit{pass@k} scores by computing their \textit{mean} over the three benchmarks. This averaging provides a single representative score for each model, facilitating a straightforward and holistic comparison of code generation capabilities across benchmarks. Models were then ranked based on the mean \textit{pass@k} scores for better comparison among their code generation capabilities. To assess the consistency of performance across benchmarks, we computed the Standard Deviation (SD) and Coefficient of Variation (CV) of \textit{pass@k} scores, capturing both absolute and relative dispersion from the mean. These measures allow us to assess how stable a model's performance is when evaluated on different tasks; lower SD and CV values indicate more consistent performance, while higher values suggest variability across benchmarks. Further to enhance interpretability, we combined SD and CV and introduced a single composite metric \textit{Stability Score (SS)}, calculated as: 
\begin{equation}
\text{SS } = \frac{1}{\left(\frac{\text{SD}}{\text{SD}_{\text{max}}} + \text{CV}\right)}
\end{equation}
This score integrates both variability measures into a unified indicator of performance stability. To ensure comparability across models, the resulting scores were normalised between 0 and 1 using \textit{min-max normalisation}, where a higher score indicates greater stability and consistency in performance. For visualisation, we used line plots to illustrate trends in SD and CV for each model in different \textit{pass@k}. Additionally, we statistically assessed whether model performance varied significantly across different groups and benchmarks by conducting a two-way ANOVA test \cite{book}. This two-way ANOVA test evaluated the influence of groups and benchmarks on the observed \textit{pass@k} scores, examining both main effects and their interaction. Following the ANOVA test, Tukey’s post-hoc test \cite{Tukey} was applied to identify statistically significant differences among the tested conditions. This post hoc analysis aligns best with our experiment, as it supports systematic pairwise comparisons in settings with relatively balanced groups, such as our evaluation setup. Together, the evaluation metrics and statistical tests provided a thorough approach to analysing model performance, enabling us to assess both accuracy and consistency.

To answer RQ2, we analysed the computational efficiency of SLMs in terms of their resource usage and inference speed. We used the same three benchmarks as in RQ1. For each model, we computed the VRAM usage and inference time across the benchmarks. VRAM usage and inference time are widely recognised as key indicators of computational efficiency, as they directly reflect the hardware resource demands and latency of model response \cite{DistilBERT}. Then we aggregate these scores by calculating their mean to understand the models overall efficiency. Besides, we statistically evaluate the models computational efficiency across different groups by conducting a one-way ANOVA test, because our analysis involved a single independent variable and its effect on computational efficiency metrics. This test evaluates the effect of groups on the resource consumption (VRAM usage) and inference speed. Following this, we applied Tukey’s post-hoc test to compare group-wise differences in resource usage. Additionally, to examine the trade-offs between model performance and efficiency, we visualised the aggregated results using a bubble chart. 

To address RQ3, we evaluated SLMs multilingual capabilities using two benchmarks explicitly designed to assess code generation capabilities across multiple programming languages. In contrast, the benchmarks we used in RQ1 and RQ2 (HumanEval, MBPP, and Mercury) primarily target only on Python and do not support multilingual evaluation. HumanEvalPack includes three different types of tasks to assess code generation capabilities in four programming languages using the \textit{pass@k} metric. To obtain a single performance measure per language, we aggregated the scores by computing the mean \textit{pass@k} across tasks. We then conducted a one-way ANOVA test, as our analysis involved a single independent variable and its effect on performance. The test results evaluate whether the observed differences in performance across languages were statistically significant or not. To complement this analysis, we used box plots to visualise the distribution of model performance by language in different mean \textit{pass@k} levels, capturing variation and outliers that indicate consistency or disparity in language capabilities.


\section{Experimental Results}
\label{Experimental Results}


In this section, we present the outcomes of our benchmark-based evaluation of 20 SLMs presented in Table \ref{tab:slm-overview}. Models are divided into three groups based on the parameter sizes to analyse any trends relative to the model size. The results are organised based on our three research questions formulated in Section \ref{Introduction}.

\subsection{SLMs Performance (RQ1)} 
\label{SLMsEffectiveness(RQ1)}

\renewcommand{\arraystretch}{1.4} 
\begin{figure*}[!b]

\begin{minipage}{\textwidth}
\captionof{table}{Code Generation Benchmark Results of 20 SLMs}
\label{tab:code_perf}
\end{minipage}

\centering
\begin{adjustbox}{max width=\textwidth}

\FloatBarrier
\begin{tabular}{|l|c|c|c|c|c|c|c|c|c|c|c|c|c|c|c|c|}
\hline
\rowcolor{lightgray}
\textbf{Model\_Parameter} &
\multicolumn{4}{c|}{\textbf{HumanEval}} &
\multicolumn{4}{c|}{\textbf{MBPP}} &
\multicolumn{4}{c|}{\textbf{Mercury}} &
\multicolumn{4}{c|}{\textbf{MEAN}} \\
\hline
\rowcolor{lightgray}
& pass@1 & pass@2 & pass@5 & pass@10 & pass@1 & pass@2 & pass@5 & pass@10 & pass@1 & pass@2 & pass@5 & pass@10 & pass@1 & pass@2 & pass@5 & pass@10 \\
\hline

\multicolumn{17}{|c|}{\rule{0pt}{2.6ex}\textbf{Group 1: SLMs parameter size $\leq$1.5B}\rule[-1.2ex]{0pt}{0pt}} \\ \hline

PolyCoder\_0.4B &
0.03 & 0.04 & 0.04 & 0.04 &
0.00 & 0.00 & 0.01 & 0.01 &
0.00 & 0.01 & 0.01 & 0.01 &
0.01 & 0.02 & 0.02 & 0.02 \\
\hline

Incoder\_1.3B &
0.09 & 0.11 & 0.13 & 0.15 &
0.09 & 0.13 & 0.18 & 0.22 &
0.01 & 0.01 & 0.02 & 0.02 &
0.06 & 0.08 & 0.11 & 0.13 \\
\hline

Phi1\_1.3B &
0.52 & 0.56 & 0.60 & 0.61 &
0.22 & 0.25 & 0.27 & 0.29 &
0.49 & 0.54 & 0.59 & 0.61 &
0.41 & 0.45 & 0.49 & 0.50 \\
\hline

DeepSeek-Coder\_1.3B &
0.32 & 0.36 & 0.42 & 0.43 &
0.44 & 0.49 & 0.54 & 0.58 &
0.54 & 0.60 & 0.66 & 0.70 &
0.43 & 0.48 & 0.54 & 0.57 \\
\hline

\rowcolor{lightgray!50}
\textbf{OpenCodeInterpreter\_1.3B} &
0.59 & 0.65 & 0.70 & 0.72 &
0.44 & 0.47 & 0.50 & 0.52 &
0.56 & 0.61 & 0.66 & 0.69 &
\textbf{0.53} & \textbf{0.58} & \textbf{0.62} & \textbf{0.64} \\
\hline

Yi-Coder\_1.5B &
0.38 & 0.42 & 0.47 & 0.50 &
0.46 & 0.51 & 0.56 & 0.59 &
0.63 & 0.68 & 0.72 & 0.75 &
0.49 & 0.54 & 0.58 & 0.61 \\
\hline

\rowcolor{lightgray!50}
\textbf{Qwen2.5-Coder\_1.5B} &
0.44 & 0.51 & 0.58 & 0.61 &
0.51 & 0.58 & 0.64 & 0.68 &
0.66 & 0.71 & 0.74 & 0.76 &
\textbf{0.54} & \textbf{0.60} & \textbf{0.65} & \textbf{0.68} \\
\hline

\multicolumn{17}{|c|}{\rule{0pt}{2.6ex}\textbf{Group 2: SLMs parameter size $>$1.5B to $\leq$3B}\rule[-1.2ex]{0pt}{0pt}} \\ \hline

CodeGemma\_2.0B &
0.07 & 0.09 & 0.12 & 0.15 &
0.31 & 0.40 & 0.48 & 0.52 &
0.47 & 0.53 & 0.58 & 0.60 &
0.28 & 0.34 & 0.39 & 0.42 \\
\hline

PolyCoder\_2.7B &
0.06 & 0.07 & 0.08 & 0.09 &
0.02 & 0.03 & 0.06 & 0.09 &
0.00 & 0.01 & 0.02 & 0.02 &
0.03 & 0.04 & 0.05 & 0.07 \\
\hline

Stable Code\_3.0B &
0.21 & 0.25 & 0.29 & 0.31 &
0.25 & 0.31 & 0.38 & 0.43 &
0.36 & 0.47 & 0.60 & 0.65 &
0.27 & 0.34 & 0.42 & 0.46 \\
\hline

StarCoder2\_3.0B &
0.32 & 0.36 & 0.42 & 0.47 &
0.41 & 0.47 & 0.53 & 0.56 &
0.54 & 0.60 & 0.64 & 0.67 &
0.42 & 0.48 & 0.53 & 0.57 \\
\hline

\rowcolor{lightgray!50}
\textbf{Qwen2.5-Coder\_3.0B} &
0.51 & 0.57 & 0.63 & 0.66 &
0.57 & 0.63 & 0.68 & 0.71 &
0.68 & 0.72 & 0.76 & 0.78 &
\textbf{0.59} & \textbf{0.64} & \textbf{0.69} & \textbf{0.72} \\
\hline

\multicolumn{17}{|c|}{\rule{0pt}{2.6ex}\textbf{Group 3: SLMs parameter size $>$3B to $\leq$10B}\rule[-1.2ex]{0pt}{0pt}} \\ \hline

Incoder\_6.7B &
0.15 & 0.18 & 0.22 & 0.24 &
0.15 & 0.21 & 0.27 & 0.31 &
0.01 & 0.01 & 0.01 & 0.02 &
0.10 & 0.13 & 0.17 & 0.19 \\
\hline

DeepSeek-Coder\_6.7B &
0.46 & 0.52 & 0.60 & 0.66 &
0.56 & 0.62 & 0.67 & 0.69 &
0.67 & 0.71 & 0.74 & 0.76 &
0.56 & 0.62 & 0.67 & 0.70 \\
\hline

\rowcolor{lightgray!50}
\textbf{OpenCodeInterpreter\_6.7B} &
0.73 & 0.76 & 0.79 & 0.80 &
0.59 & 0.63 & 0.66 & 0.67 &
0.70 & 0.72 & 0.74 & 0.75 &
\textbf{0.67} & \textbf{0.70} & \textbf{0.73} & \textbf{0.74} \\
\hline

CodeLlama\_7.0B &
0.30 & 0.34 & 0.39 & 0.42 &
0.37 & 0.44 & 0.49 & 0.53 &
0.45 & 0.54 & 0.64 & 0.71 &
0.37 & 0.44 & 0.51 & 0.55 \\
\hline

StarCoder2\_7.0B &
0.35 & 0.40 & 0.46 & 0.50 &
0.45 & 0.51 & 0.56 & 0.59 &
0.60 & 0.65 & 0.69 & 0.71 &
0.47 & 0.52 & 0.57 & 0.60 \\
\hline

CodeGemma\_7.0B &
0.42 & 0.52 & 0.65 & 0.74 &
0.46 & 0.57 & 0.67 & 0.72 &
0.60 & 0.68 & 0.73 & 0.75 &
0.49 & 0.59 & 0.68 & 0.74 \\
\hline

\rowcolor{lightgray!50}
\textbf{Qwen2.5-Coder\_7.0B} &
0.62 & 0.69 & 0.74 & 0.77 &
0.63 & 0.68 & 0.73 & 0.76 &
0.70 & 0.73 & 0.75 & 0.77 &
\textbf{0.65} & \textbf{0.70} & \textbf{0.74} & \textbf{0.77} \\
\hline

Yi-Coder\_9.0B &
0.51 & 0.58 & 0.66 & 0.70 &
0.57 & 0.64 & 0.70 & 0.73 &
0.69 & 0.73 & 0.77 & 0.77 &
0.59 & 0.65 & 0.71 & 0.73 \\
\hline

\end{tabular}
\end{adjustbox}

\end{figure*}

\renewcommand{\arraystretch}{1.3} 

\begin{table}[!H] 
\centering
\caption{Ranking of SLMs based on MEAN \textit{pass@k} scores (Lower is Better)}
\label{tab:group-best}

\resizebox{\columnwidth}{!}{
\begin{tabular}{|l|c|c|c|c|}
\hline
\rowcolor{lightgray}
\textbf{Model\_Parameter} & \textbf{pass@1} & \textbf{pass@2} & \textbf{pass@5} & \textbf{pass@10} \\
\hline

\multicolumn{5}{|c|}{\textbf{Group 1: SLMs parameter size  $\leq$ 1.5B}} \\ \hline
PolyCoder\_0.4B & 20 & 20 & 20 & 20 \\ 
\hline
Incoder\_1.3B & 18 & 18 & 18 & 18 \\ 
\hline
Phi1\_1.3B & 13 & 13 & 14 & 14 \\ 
\hline
DeepSeek-Coder\_1.3B & 11 & 11 & 11 & 11 \\ 
\hline
\rowcolor{lightgray!50}
\textbf{OpenCodeInterpreter\_1.3B} & \textbf{7} & \textbf{8} & \textbf{8} & \textbf{8} \\ 
\hline
Yi-Coder\_1.5B & 9 & 9 & 9 & 9 \\ 
\hline
\rowcolor{lightgray!50}
\textbf{Qwen2.5-Coder\_1.5B} & \textbf{6} & \textbf{6} & \textbf{7} & \textbf{7} \\
\hline

\multicolumn{5}{|c|}{\textbf{Group 2: SLMs parameter size  $>$1.5B to $\leq$3B}} \\ \hline
CodeGemma\_2.0B & 15 & 16 & 16 & 16 \\ 
\hline
PolyCoder\_2.7B & 19 & 19 & 19 & 19 \\ 
\hline
Stable Code\_3.0B & 16 & 15 & 15 & 15 \\ 
\hline
StarCoder2\_3.0B & 12 & 12 & 12 & 12 \\ 
\hline
\rowcolor{lightgray!50}
\textbf{Qwen2.5-Coder\_3.0B} & \textbf{4} & \textbf{4} & \textbf{4} & \textbf{5} \\
\hline

\multicolumn{5}{|c|}{\textbf{Group 3: SLMs parameter size  $>$3B to $\leq$10B}} \\ \hline
Incoder\_6.7B & 17 & 17 & 17 & 17 \\ 
\hline
DeepSeek-Coder\_6.7B & 5 & 5 & 6 & 6 \\ 
\hline
\rowcolor{lightgray!50}
\textbf{OpenCodeInterpreter\_6.7B} & \textbf{1} & \textbf{1} & 2 & 2 \\ 
\hline
CodeLlama\_7.0B & 14 & 14 & 13 & 13 \\ 
\hline
StarCoder2\_7.0B & 10 & 10 & 10 & 10 \\ 
\hline
CodeGemma\_7.0B & 8 & 7 & 5 & 3 \\ 
\hline
\rowcolor{lightgray!50}
\textbf{Qwen2.5-Coder\_7.0B} & \textbf{2} & \textbf{2} & \textbf{1} & \textbf{1} \\ 
\hline
Yi-Coder\_9.0B & 3 & 3 & 3 & 4 \\ 
\hline

\end{tabular}
}
\label{tab:code_perf_rank}
\end{table}

We evaluate the SLMs performance on two main aspects: (i) functional correctness of their generated outputs and (ii) stability of generating functionally correct solutions across three benchmarks, which are HumanEval, MBPP, and Mercury. Functional correctness is measured using the \textit{pass@k} metric. This metric indicates the likelihood that at least one of the top-k generated outputs passes all the test cases. It is well-suited for evaluating code generation models, as it reflects their ability to produce correct outputs in stochastic generation scenarios. Table~\ref{tab:code_perf} reports the \textit{pass@k} scores (k = 1, 2, 5, 10) for each model on these benchmarks, along with their mean scores. These results reflect the functional correctness of the models and provide a basis for comparing their overall performance. Besides, Table \ref{tab:code_perf_rank} presents the \textit{rank@k} positions of the models based on their mean \textit{pass@k} scores. This ranking reveals several noteworthy pieces of information about model performance across different \textit{pass@k} levels. 

Across the benchmarks, models from Group 3 where parameter size $>$3B and $\leq$10B, generally achieve the highest \textit{pass@k} scores and occupy the top positions in the rankings. For instance, OpenCodeInterpreter 6.7B, Qwen2.5-Coder 7B, and Yi-Coder 9.0B consistently rank among the top four across all \textit{pass@k} values, with mean \textit{pass@1} scores ranging from 0.59 to 0.67, outperforming the smaller SLMs from Groups 1 and 2. This trend persists across \textit{pass@2}, \textit{pass@5}, and \textit{pass@10}, reflecting the advantage of larger model capacity in producing correct code samples. However, the results also show that several mid-sized and small models perform competitively, particularly within Group 1 and Group 2. Qwen2.5-Coder 3.0B from Group 2 achieves a mean \textit{pass@1} score of 0.59, placing it among the top five models and performing comparably to larger Group 3 models. Similarly, Group 1 models such as OpenCodeInterpreter 1.3B and Qwen2.5-Coder 1.5B rank within the top eight with a mean \textit{pass@1} score of 0.53 and 0.54, outperforming several larger SLMs (e.g., CodeLlama 7B and StarCoder2 7B with a mean \textit{pass@1} score of 0.37 and 0.47). 

\renewcommand{\arraystretch}{1.5} 

\begin{table*}[!H]
\caption{Standard Deviation (SD), Coefficient of Variation (COV) and Normalised Stability Score (NSS) of 20 SLMs in different \textit{pass@k}}
\centering

\resizebox{\textwidth}{!}{%
{
\begin{tabular}{|l|
    c|c|c|
    c|c|c|
    c|c|c|
    c|c|c|}
\hline
\rowcolor{lightgray}
\textbf{Model\_Parameter} &
\multicolumn{3}{c|}{\textbf{pass@1}} &
\multicolumn{3}{c|}{\textbf{pass@2}} &
\multicolumn{3}{c|}{\textbf{pass@5}} &
\multicolumn{3}{c|}{\textbf{pass@10}} \\
\hline
\rowcolor{lightgray}
& {SD} & {COV} & {NSS} &
{SD} & {COV} & {NSS} &
{SD} & {COV} & {NSS} &
{SD} & {COV} & {NSS} \\
\hline

\multicolumn{13}{|c|}{\rule{0pt}{2.6ex}\textbf{Group 1: SLMs parameter size $\leq$1.5B}\rule[-1.2ex]{0pt}{0pt}} \\ \hline

PolyCoder\_0.4B &
0.02 & 1.73 & 0.00 &
0.02 & 1.25 & 0.02 &
0.02 & 0.87 & 0.03 &
0.02 & 0.65 & 0.02 \\
\hline

Incoder\_1.3B &
0.05 & 0.73 & 0.17 &
0.06 & 0.77 & 0.06 &
0.08 & 0.74 & 0.02 &
0.10 & 0.78 & 0.01 \\
\hline

Phi1\_1.3B &
0.17 & 0.40 & 0.09 &
0.17 & 0.39 & 0.05 &
0.19 & 0.39 & 0.01 &
0.18 & 0.37 & 0.01 \\
\hline

DeepSeek-Coder\_1.3B &
0.11 & 0.25 & 0.23 &
0.12 & 0.25 & 0.12 &
0.12 & 0.22 & 0.04 &
0.14 & 0.24 & 0.02 \\
\hline

\rowcolor{lightgray!50}
\textbf{OpenCodeInterpreter\_1.3B} &
\textbf{0.08} & \textbf{0.15} & \textbf{0.43} &
\textbf{0.09} & \textbf{0.16} & \textbf{0.19} &
\textbf{0.11} & \textbf{0.17} & \textbf{0.06} &
\textbf{0.11} & \textbf{0.17} & \textbf{0.03} \\
\hline

Yi-Coder\_1.5B &
0.13 & 0.26 & 0.19 &
0.13 & 0.25 & 0.10 &
0.13 & 0.22 & 0.04 &
0.13 & 0.21 & 0.02 \\
\hline

\rowcolor{lightgray!50}
\textbf{Qwen2.5-Coder\_1.5B} &
\textbf{0.11} & \textbf{0.21} & \textbf{0.25} &
\textbf{0.10} & \textbf{0.17} & \textbf{0.17} &
\textbf{0.08} & \textbf{0.12} & \textbf{0.09} &
\textbf{0.08} & \textbf{0.11} & \textbf{0.06} \\
\hline


\multicolumn{13}{|c|}{\rule{0pt}{2.6ex}\textbf{Group 2: SLMs parameter size $>$1.5B to $\leq$3B}\rule[-1.2ex]{0pt}{0pt}} \\ \hline

CodeGemma\_2.0B &
0.20 & 0.71 & 0.01 &
0.23 & 0.66 & 0.00 &
0.24 & 0.62 & 0.00 &
0.24 & 0.57 & 0.00 \\
\hline

PolyCoder\_2.7B &
0.03 & 1.15 & 0.07 &
0.03 & 0.83 & 0.07 &
0.03 & 0.57 & 0.05 &
0.04 & 0.61 & 0.02 \\
\hline

Stable Code\_3.0B &
0.08 & 0.28 & 0.32 &
0.11 & 0.33 & 0.10 &
0.16 & 0.38 & 0.02 &
0.17 & 0.37 & 0.01 \\
\hline

StarCoder2\_3.0B &
0.11 & 0.26 & 0.23 &
0.12 & 0.25 & 0.12 &
0.11 & 0.21 & 0.05 &
0.10 & 0.18 & 0.03 \\
\hline

\rowcolor{lightgray!50}
\textbf{Qwen2.5-Coder\_3.0B} &
\textbf{0.09} & \textbf{0.15} & \textbf{0.40} &
\textbf{0.08} & \textbf{0.12} & \textbf{0.28} &
\textbf{0.07} & \textbf{0.10} & \textbf{0.12} &
\textbf{0.06} & \textbf{0.08} & \textbf{0.08} \\
\hline


\multicolumn{13}{|c|}{\rule{0pt}{2.6ex}\textbf{Group 3: SLMs parameter size $>$3B to $\leq$10B}\rule[-1.2ex]{0pt}{0pt}} \\ \hline

Incoder\_6.7B &
0.08 & 0.78 & 0.10 &
0.11 & 0.81 & 0.03 &
0.14 & 0.83 & 0.01 &
0.15 & 0.80 & 0.00 \\
\hline

DeepSeek-Coder\_6.7B &
0.11 & 0.19 & 0.29 &
0.10 & 0.15 & 0.19 &
0.07 & 0.10 & 0.11 &
0.05 & 0.07 & 0.09 \\
\hline

OpenCodeInterpreter\_6.7B &
0.07 & 0.11 & 0.52 &
0.07 & 0.09 & 0.34 &
0.07 & 0.09 & 0.12 &
0.07 & 0.09 & 0.07 \\
\hline

CodeLlama\_7.0B &
0.08 & 0.20 & 0.40 &
0.10 & 0.23 & 0.15 &
0.13 & 0.25 & 0.04 &
0.15 & 0.26 & 0.02 \\
\hline

StarCoder2\_7.0B &
0.13 & 0.27 & 0.19 &
0.13 & 0.24 & 0.11 &
0.12 & 0.20 & 0.05 &
0.11 & 0.18 & 0.03 \\
\hline

CodeGemma\_7.0B &
0.09 & 0.19 & 0.32 &
0.08 & 0.14 & 0.24 &
0.04 & 0.06 & 0.21 &
0.02 & 0.02 & 0.36 \\
\hline

\rowcolor{lightgray!50}
\textbf{Qwen2.5-Coder\_7.0B} &
\textbf{0.04} & \textbf{0.07} & \textbf{1.00} &
\textbf{0.03} & \textbf{0.04} & \textbf{1.00} &
\textbf{0.01} & \textbf{0.01} & \textbf{1.00} &
\textbf{0.01} & \textbf{0.01} & \textbf{1.00} \\
\hline

Yi-Coder\_9.0B &
0.09 & 0.16 & 0.37 &
0.08 & 0.12 & 0.28 &
0.06 & 0.08 & 0.15 &
0.04 & 0.05 & 0.15 \\
\hline

\end{tabular}
}
}
\label{tab:stability_scores_normalized}
\end{table*}

Beyond these trends, additional patterns emerge that further highlight the diversity of SLM behaviour. Models such as DeepSeek-Coder 6.7B show consistent performance across all benchmarks, with balanced improvements in \textit{pass@1} to \textit{pass@10}, reflecting robust generalisation capabilities. CodeGemma 7B exhibits substantial gains from pass@1 to pass@10, indicating that it benefits greatly from sampling diversity, as it generates multiple near-correct answer variants. In contrast, models like StableCode 3B and CodeLlama 7B show uneven behaviour across benchmarks. They show relatively weaker performance on HumanEval but noticeably stronger on Mercury, suggesting that some model families may be particularly suited to algorithmic or dataflow-heavy problems rather than natural-language-driven coding tasks. Among smaller models, Phi-1.3B displays strong performance on HumanEval (pass@1 = 0.52) but much weaker results on MBPP (pass@1 = 0.22), likely reflecting differences in training data and domain specialisation. At the lower end, PolyCoder (0.4B and 2.7B) and Incoder (1.3B and 6.7B) show consistently weak performance, with \textit{pass@1} values below 0.10 across benchmarks. This pattern remains stable across \textit{pass@2}, \textit{pass@5}, and \textit{pass@10}, indicating that their lower performance is not simply a top-1 effect. These findings indicate that parameter count alone does not fully determine performance; model architecture and training quality also play significant roles.

\begin{table}[!htbp]
\centering
\caption{Ranking of SLMs based on Stability Score (Lower Rank = Better)}
\label{tab:stability-best}

\resizebox{\columnwidth}{!}{
\begin{tabular}{|l|c|c|c|c|}
\hline
\rowcolor{lightgray}
\textbf{Model\_Parameter} &
\textbf{pass@1} &
\textbf{pass@2} &
\textbf{pass@5} &
\textbf{pass@10} \\
\hline

\multicolumn{5}{|c|}{\textbf{Group 1: SLMs parameter size  $\leq$ 1.5B}} \\ \hline
PolyCoder\_0.4B & 20 & 19 & 15 & 11 \\ 
\hline
Incoder\_1.3B & 15 & 16 & 17 & 18 \\ 
\hline
Phi1\_1.3B & 17 & 17 & 18 & 17 \\ 
\hline
DeepSeek-Coder\_1.3B & 11 & 10 & 12 & 14 \\ 
\hline
\rowcolor{lightgray!50}
\textbf{OpenCodeInterpreter\_1.3B }& \textbf{3} & \textbf{7} & \textbf{8} & \textbf{10} \\ 
\hline
Yi-Coder\_1.5B & 14 & 13 & 13 & 12 \\ 
\hline
Qwen2.5-Coder\_1.5B & 10 & 8 & 7 & 7 \\ 
\hline

\multicolumn{5}{|c|}{\textbf{Group 2: SLMs parameter size  $>$1.5B to $\leq$3B}} \\ \hline
CodeGemma\_2.0B & 19 & 20 & 20 & 20 \\ 
\hline
PolyCoder\_2.7B & 18 & 15 & 11 & 13 \\ 
\hline
Stable Code\_3.0B & 8 & 14 & 16 & 16 \\ 
\hline
StarCoder2\_3.0B & 12 & 11 & 9 & 8 \\ 
\hline
\rowcolor{lightgray!50}
\textbf{Qwen2.5-Coder\_3.0B} & \textbf{5} & \textbf{4} & \textbf{5} & \textbf{5} \\ 
\hline

\multicolumn{5}{|c|}{\textbf{Group 3: SLMs parameter size  $>$3B to $\leq$10B}} \\ \hline

Incoder\_6.7B & 16 & 18 & 19 & 19 \\ 
\hline
DeepSeek-Coder\_6.7B & 9 & 6 & 6 & 4 \\ 
\hline
\rowcolor{lightgray!50}
\textbf{OpenCodeInterpreter\_6.7B} & \textbf{2} & \textbf{2} & \textbf{4} & \textbf{6} \\ 
\hline
CodeLlama\_7.0B & 4 & 9 & 14 & 15 \\ 
\hline
StarCoder2\_7.0B & 13 & 12 & 10 & 9 \\ 
\hline
CodeGemma\_7.0B & 7 & 5 & 2 & 2 \\ 
\hline
\rowcolor{lightgray!50}
\textbf{Qwen2.5-Coder\_7.0B} & \textbf{1} & \textbf{1} & \textbf{1} & \textbf{1} \\ 
\hline
Yi-Coder\_9.0B & 6 & 3 & 3 & 3 \\ 
\hline

\end{tabular}
}
\label{tab:code_stab_rank}
\end{table}

\begin{figure*}[!t]
    \centering
    \setlength{\abovecaptionskip}{2pt}
    \setlength{\belowcaptionskip}{2pt}
    \setlength{\intextsep}{0pt}
    \setlength{\textfloatsep}{5pt}

    \subfloat[pass@1]{%
        \includegraphics[width=0.80\linewidth]{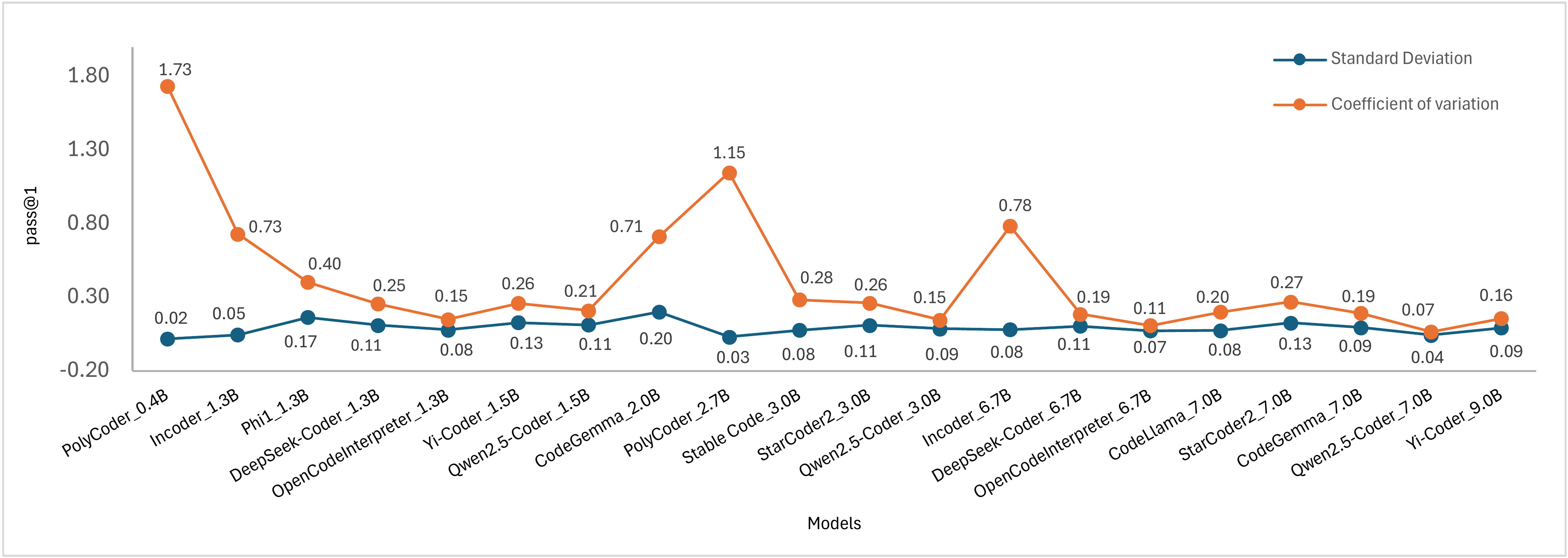}
    }\vspace{-8pt}\par

    \subfloat[pass@2]{%
        \includegraphics[width=0.80\linewidth]{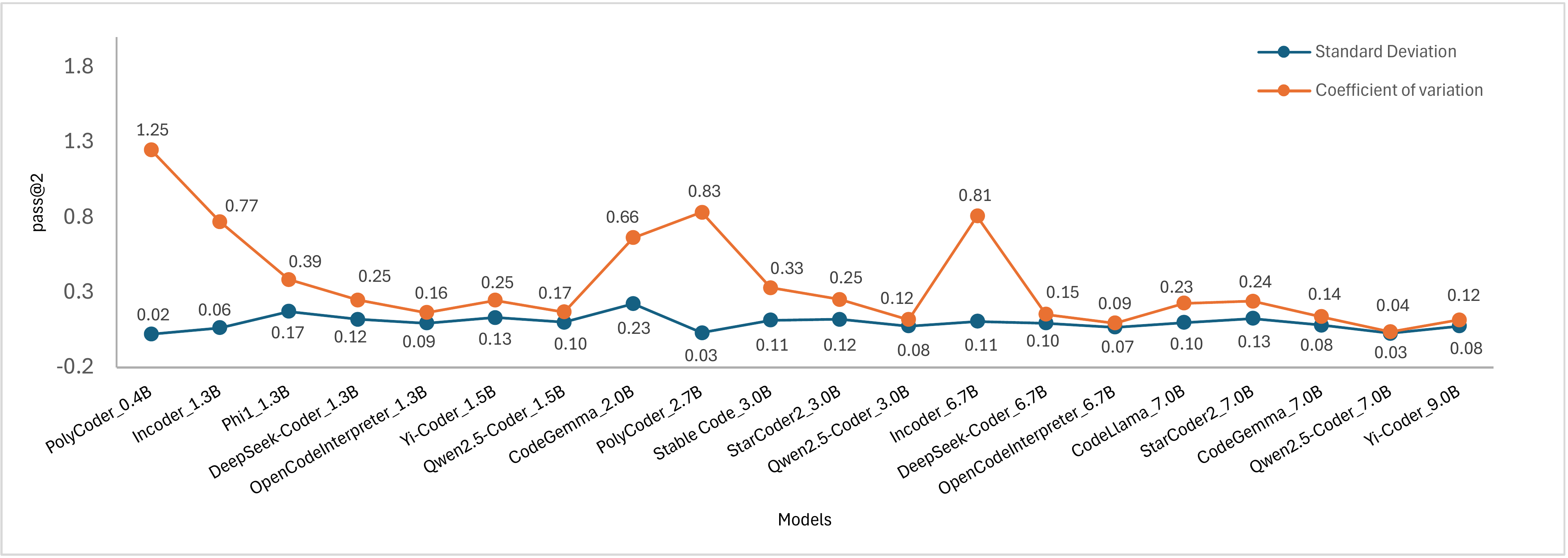}
    }\vspace{-8pt}\par

    \subfloat[pass@5]{%
        \includegraphics[width=0.80\linewidth]{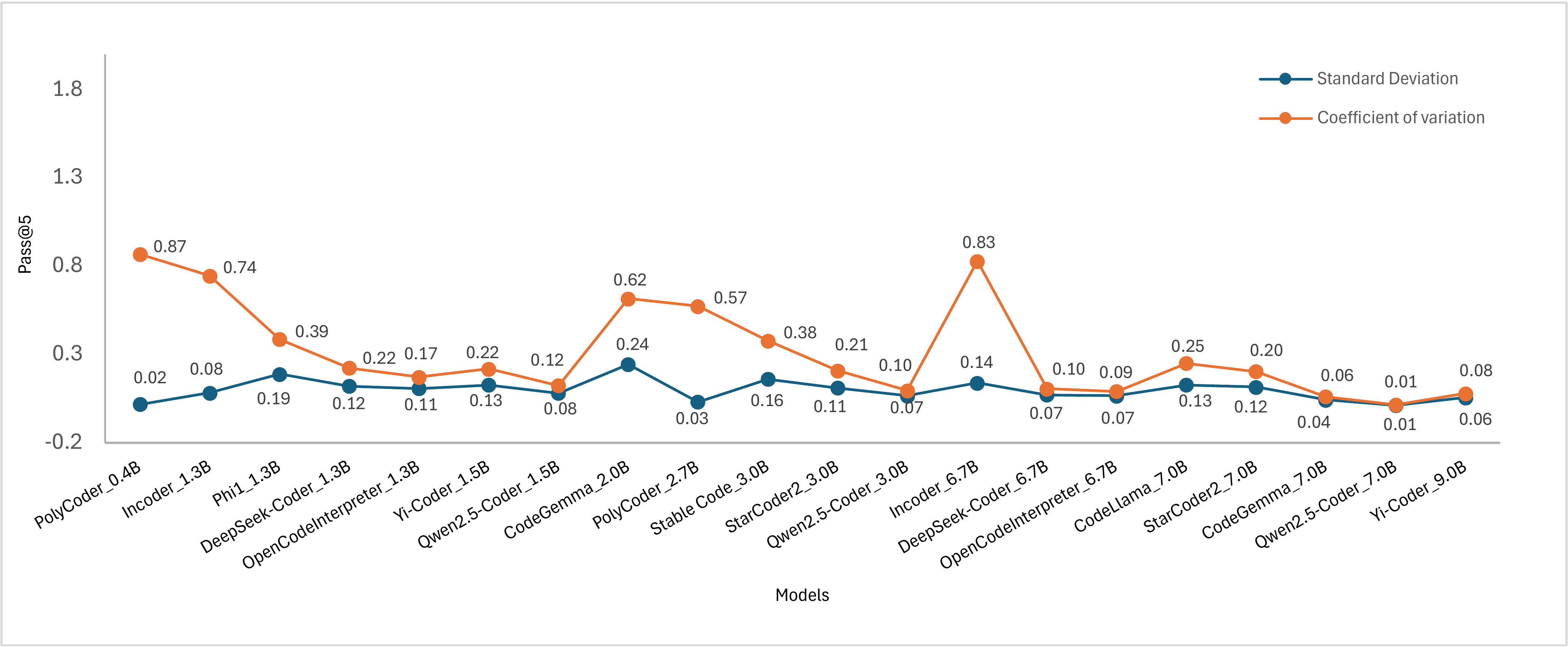}
    }\vspace{-8pt}\par

    \subfloat[pass@10]{%
        \includegraphics[width=0.80\linewidth]{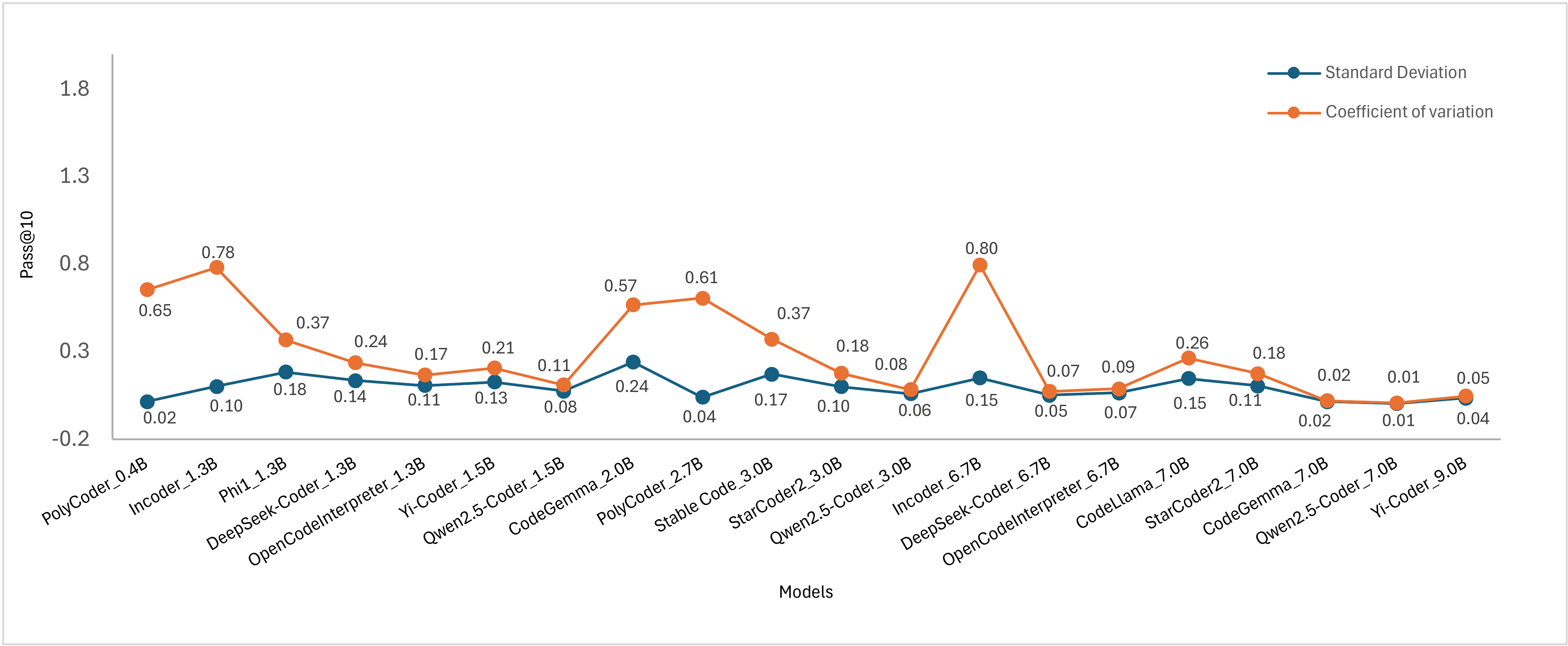}
    }\vspace{-4pt}\par

    \caption{Standard deviation(SD) and coefficient of variation(COV) of the models. Subplots show the scores for (a) \textit{pass@1}, (b) \textit{pass@2}, (c) \textit{pass@5}, and (d) \textit{pass@10}. A smaller gap between SD and COV indicates stable performance across benchmarks.}
    \label{fig:rq1_2}
\end{figure*}

Besides functional correctness, it is important to understand how consistently each model performs, which we examine through a stability score introduced in section \ref{sub:data_analysis}. Table~\ref{tab:stability_scores_normalized} presents the normalised stability scores of all the selected SLMs across different \textit{pass@k} levels, alongside their standard deviation and coefficient of variation. These metrics reflect the stability of each model's performance across the three benchmarks. Figure~\ref{fig:rq1_2} offer a visual breakdown of these metrics. It shows the standard deviation and coefficient of variation for each model at different \textit{pass@k} levels. In the figure, the models are plotted along the x-axis, while the values of standard deviation and coefficient of variation are shown on the y-axis. This visualisation highlights how much each model's performance fluctuates from its mean across benchmarks in different \textit{pass@k} levels. Models with lower standard deviation and coefficient of variation, along with a smaller gap between the two, are more stable in their performance across benchmarks. Table~\ref{tab:code_stab_rank} showcases the \textit{rank@k} position of the models based on their normalised stability scores across all \textit{pass@k} levels. These results across the table and figures highlight the following insights about SLMs stability.

Overall, Group 3 models show the highest stability. For example, Qwen2.5-Coder 7B achieves the maximum Stability Score of 1.00 across all \textit{pass@k} levels. Other Group 3 models, such as OpenCodeInterpreter 6.7B, Yi-Coder 9.0B, also obtain decent Stability Scores of 0.52 and 0.37, respectively, at the \textit{pass@1} level. In contrast, Groups 1 and 2 exhibit wider variation. While Qwen2.5-Coder 3.0B (0.40 at \textit{pass@1}) and OpenCodeInterpreter 1.3B (0.43 at \textit{pass@1}) rank within the top ten across multiple \textit{pass@k} levels,  models like PolyCoder 0.4B and Incoder 1.3B obtain some of the lowest Stability Scores, with values as low as 0.00 and 0.17.

However, the stability results also reveal several nuanced patterns that are not captured by functional correctness alone. Many Group 3 models become more stable as "k(e.g., 1,2,5,10)" increases; for instance, CodeGemma 7B and StarCoder2 7B show noticeably lower variability at pass@10 than at pass@1, suggesting that larger models benefit from multi-sample generation where randomness is averaged out. At the same time, stability does not always correlate with performance. Some models with competitive pass@k scores exhibit comparatively high fluctuation—StarCoder2 7B, for example, achieves a respectable mean pass@1 of 0.47 but has a modest Stability Score of 0.19, indicating inconsistent behaviour across benchmarks. Conversely, certain smaller models demonstrate both strong performance and high stability. OpenCodeInterpreter 1.3B, despite its size, attains a mean pass@1 of 0.53 alongside an NSS of 0.43, outperforming several larger models. A notable pattern emerges within model families as well: the Qwen family demonstrates consistently high stability relative to size, particularly Qwen2.5-Coder 3B and Qwen2.5-Coder 7B, which maintain low variation and strong NSS values across all \textit{pass@k} levels. This suggests that architectural and training pipeline design can play a substantial role in promoting stability, independent of parameter count.

The observed performance of the models was further validated statistically by conducting a two-way ANOVA test. This test assesses any influence on performance by (i) the size of the model (group) and (ii) the benchmark used, as well as whether there is any interaction between these factors. As reported in Table \ref{tab:anova2w_passk}, the results show that the model group has a statistically significant effect on performance across all \textit{pass@k} levels (p \(<\) 0.05). In contrast, the choice of benchmark and the interaction between model group and benchmark were not statistically significant. 

To further understand where the differences lie, Table \ref{tab:tukey_posthoc} reports the Tukey’s HSD post-hoc test results. The results reveal that Group 3 models (largest) significantly outperform Group 2 models (mid-sized) at \textit{pass@1}, while the differences at \textit{pass@2}, \textit{pass@5}, and \textit{pass@10} are marginally significant or lie near the statistical threshold. However, the performance gap between Group 1 (smallest) and Group 3 is not statistically significant in \textit{pass@1,2, and 5} but significant in \textit{pass@10}.  Similarly, the performance gap between groups 1 and 2 is not statistically significant. However, their mean difference is negative in all \textit{pass@k}.

\renewcommand{\arraystretch}{1.5} 

\begin{table}[!h]
\caption{Two-way ANOVA Results for the Effects of Model Group and Benchmark on \textit{pass@k} Scores}
\centering
\resizebox{\columnwidth}{!}{
\begin{tabular}{|c|l|c|c|c|}
\hline
\rowcolor{lightgray}
\textbf{Metric} & \textbf{Source} & \textbf{F-statistic} & \textbf{p-value} & \makecell{\textbf{Significant}} \\
\hline
\multirow{3}{*}{pass@1} 
& Group             & 3.63 & 0.0337 & \textbf{Yes} \\
& Benchmark         & 1.59 & 0.2137 & No \\
& Group:Benchmark   & 0.12 & 0.9739 & No \\
\hline
\multirow{3}{*}{pass@2} 
& Group             & 3.68 & 0.0323 & \textbf{Yes} \\
& Benchmark         & 1.34 & 0.2712 & No \\
& Group:Benchmark   & 0.17 & 0.9503 & No \\
\hline
\multirow{3}{*}{pass@5} 
& Group             & 3.67 & 0.0323 & \textbf{Yes} \\
& Benchmark         & 0.99 & 0.3785 & No \\
& Group:Benchmark   & 0.23 & 0.9192 & No \\
\hline
\multirow{3}{*}{pass@10} 
& Group             & 3.73 & 0.0307 & \textbf{Yes} \\
& Benchmark         & 0.80 & 0.4530 & No \\
& Group:Benchmark   & 0.23 & 0.9204 & No \\
\hline
\end{tabular}
}
\label{tab:anova2w_passk}
\end{table}

\renewcommand{\arraystretch}{1.5} 

\begin{table}[!h]
\caption{Pairwise Comparison of Model Groups on \textit{pass@k} Scores Using Tukey’s HSD Test}
\centering
\resizebox{\columnwidth}{!}{%
\begin{tabular}{|c|c|c|c|c|c|}
\hline
\rowcolor{lightgray}
\textbf{Metric} & \textbf{Candidate 1} & \textbf{Candidate 2} & \textbf{Mean Diff} & \textbf{p-adj} & \textbf{Significant} \\
\hline
\multirow{3}{*}{pass@1}
& 1 & 2 & -0.0347 & 0.8779 & No  \\
& 1 & 3 &  0.1358 & 0.0874 & No  \\
& 2 & 3 &  0.1705 & 0.0442 & \textbf{Yes} \\
\hline
\multirow{3}{*}{pass@2}
& 1 & 2 & -0.0250 & 0.9412 & No  \\
& 1 & 3 &  0.1518 & 0.0677 & No  \\
& 2 & 3 &  0.1768 & 0.0501 & No (borderline)  \\
\hline
\multirow{3}{*}{pass@5}
& 1 & 2 & -0.0125 & 0.9864 & No  \\
& 1 & 3 &  0.1666 & 0.0531 & No  \\
& 2 & 3 &  0.1791 & 0.0607 & No (marginal) \\
\hline
\multirow{3}{*}{pass@10}
& 1 & 2 & -0.0050 & 0.9979 & No \\
& 1 & 3 &  0.1755 & 0.0444 & \textbf{Yes}  \\
& 2 & 3 &  0.1806 & 0.0647 & No (marginal) \\
\hline
\end{tabular}
}
\label{tab:tukey_posthoc}
\end{table}

\begin{tcolorbox}[colback=gray!9, colframe=black, title=Key Findings from RQ1, left=0pt, right=0pt, boxsep=5pt]
\begin{itemize}
    \item The size of the model has a clear effect on its performance. In contrast, the choice of benchmark does not make a significant difference. This indicates that model performance remains relatively stable across various coding tasks.
    \item Larger SLMs perform better than smaller ones in both correctness and stability. They usually rank at the top across all \textit{pass@k} levels. However, some small and mid-sized SLMs also perform well. This indicates that strong results are possible even with fewer parameters.
\end{itemize}
\end{tcolorbox}

\subsection{Performance Trade-Offs (RQ2)}
\label{PerformanceTrade-Offs(RQ2)}

\renewcommand{\arraystretch}{1.3} 

\begin{table*}[!H]
\caption{Model-wise Summary of mean \textit{pass@k} Scores, VRAM Usage, and Inference Time}
\centering
\resizebox{\textwidth}{!}{%
\begin{tabular}{|l|c|c|c|c|c|c|}
\hline
\rowcolor{lightgray}
\textbf{Model\_Parameter} & 
\textbf{Mean\_pass@1} & \textbf{Mean\_pass@2} & 
\textbf{Mean\_pass@5} & \textbf{Mean\_pass@10} & 
\textbf{Avg VRAM (GB)} & \textbf{Avg Time (Sec)} \\
\hline

\multicolumn{7}{|c|}{\rule{0pt}{2.6ex}\textbf{Group 1: SLMs parameter size $\leq$1.5B}\rule[-1.2ex]{0pt}{0pt}} \\
\hline

PolyCoder\_0.4B &
{0.01} & {0.02} & {0.02} & {0.02} &
{5.03} & {2.49} \\
\hline

Incoder\_1.3B &
{0.06} & {0.08} & {0.11} & {0.13} &
{13.88} & {3.06} \\
\hline

Phi1\_1.3B &
{0.41} & {0.45} & {0.49} & {0.50} &
{8.45} & {1.28} \\
\hline

DeepSeek-Coder\_1.3B &
{0.43} & {0.48} & {0.54} & {0.57} &
{12.12} & {1.19} \\
\hline

\rowcolor{lightgray!50}
\textbf{OpenCodeInterpreter\_1.3B} &
\textbf{{0.53}} & 
\textbf{{0.58}} &
\textbf{{0.62}} & 
\textbf{{0.64}} &
\textbf{{11.97}} & 
\textbf{{1.06}} \\
\hline

Yi-Coder\_1.5B &
{0.49} & {0.54} & {0.58} & {0.61} &
{11.99} & {0.80} \\
\hline

\rowcolor{lightgray!50}
\textbf{Qwen2.5-Coder\_1.5B} &
\textbf{{0.54}} & \textbf{{0.60}} &
\textbf{{0.65}} & \textbf{{0.68}} &
\textbf{{6.55}} & 
\textbf{{1.39}} \\
\hline

\multicolumn{7}{|c|}{\rule{0pt}{2.6ex}\textbf{Group 2: SLMs parameter size $>$1.5B to $\leq$3B}\rule[-1.2ex]{0pt}{0pt}} \\
\hline

CodeGemma\_2.0B &
{0.28} & {0.34} & {0.39} & {0.42} &
{8.70} & {1.78} \\
\hline

PolyCoder\_2.7B &
{0.03} & {0.04} & {0.05} & {0.07} &
{14.15} & {5.58} \\
\hline

Stable Code\_3.0B &
{0.27} & {0.34} & {0.42} & {0.46} &
{17.57} & {1.64} \\
\hline

\rowcolor{lightgray!50}
\textbf{StarCoder2\_3.0B} &
\textbf{{0.42}} & \textbf{{0.48}} &
\textbf{{0.53}} & \textbf{{0.57}} &
{7.99} & {1.03} \\
\hline

\rowcolor{lightgray!50}
\textbf{Qwen2.5-Coder\_3.0B} &
\textbf{{0.59}} & \textbf{{0.64}} &
\textbf{{0.69}} & \textbf{{0.72}} &
{10.77} & {1.69} \\
\hline

\multicolumn{7}{|c|}{\rule{0pt}{2.6ex}\textbf{Group 3: SLMs parameter size $>$3B to $\leq$10B}\rule[-1.2ex]{0pt}{0pt}} \\
\hline

Incoder\_6.7B &
{0.10} & {0.13} & {0.17} & {0.19} &
{23.53} & {7.74} \\
\hline

DeepSeek-Coder\_6.7B &
{0.56} & {0.62} & {0.67} & {0.70} &
{23.30} & {3.34} \\
\hline

\rowcolor{lightgray!50}
\textbf{OpenCodeInterpreter\_6.7B} &
\textbf{{0.67}} & \textbf{{0.70}} &
\textbf{{0.73}} & \textbf{{0.74}} &
{23.32} & {2.67} \\
\hline

CodeLlama\_7.0B &
{0.37} & {0.44} & {0.51} & {0.55} &
{23.58} & {2.11} \\
\hline

\rowcolor{lightgray!50}
\textbf{StarCoder2\_7.0B} &
\textbf{{0.47}} & \textbf{{0.52}} &
\textbf{{0.57}} & \textbf{{0.60}} &
{17.00} & {2.92} \\
\hline

CodeGemma\_7.0B &
{0.49} & {0.59} & {0.68} & {0.74} &
{23.38} & {1.67} \\
\hline

\rowcolor{lightgray!50}
\textbf{Qwen2.5-Coder\_7.0B} &
\textbf{{0.65}} & \textbf{{0.70}} &
\textbf{{0.74}} & \textbf{{0.77}} &
{23.70} & {0.97} \\
\hline

Yi-Coder\_9.0B &
{0.59} & {0.65} & 
{0.71} & {0.73} &
{20.33} &
{3.99} \\
\hline

\end{tabular}}
\label{tab:passk-vram-inference}
\end{table*}

\begin{figure*}[!h]
\centering
\includegraphics[width=\textwidth]{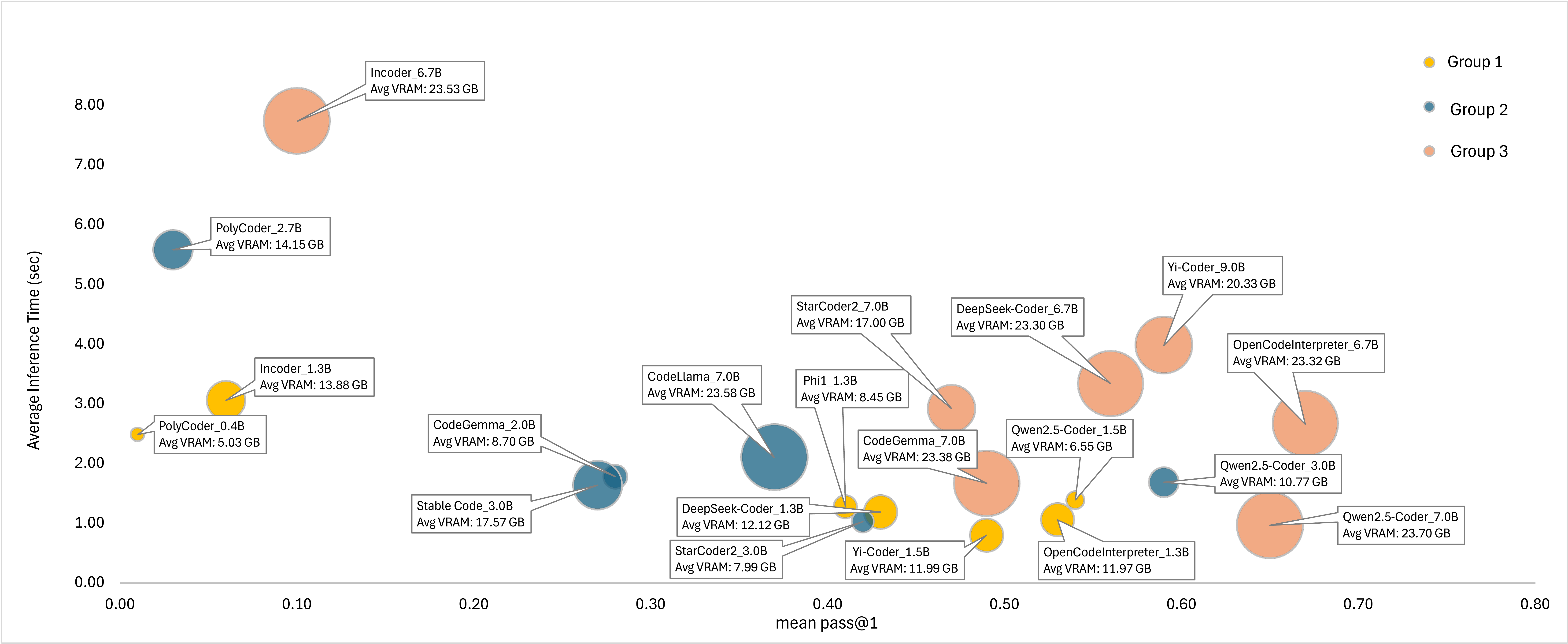}
\caption{Performance–Efficiency Trade-offs: pass@1 vs. Inference Time with VRAM Usage}
\label{fig:rq2_1}
\end{figure*}

The performance trade-offs of a model refer to the balance between a model’s ability to generate functionally correct code and the computational resources it consumes during inference. Table~\ref{tab:passk-vram-inference} presents each model’s average \textit{pass@k} scores, VRAM usage, and inference time, all measured across the same three benchmarks used in RQ1: HumanEval, MBPP, and Mercury. Figure \ref{fig:rq2_1} provides a visual representation of the performance–efficiency trade-offs for selected SLMs by plotting their mean \textit{pass@1} scores (x-axis) against their mean inference time (y-axis). Each model is represented as a bubble, where the size of the bubble corresponds to its mean VRAM usage in GB. Larger bubbles indicate higher memory requirements. This visualisation allows for a quick comparison of how models balance accuracy and resource consumption. Models located in the bottom-right quadrant of the plot, with smaller bubbles, are considered ideal, as they combine high accuracy, fast inference, and low memory usage. From the results, we can see that some models combine relatively high accuracy with faster response and low resource usage. For example, Qwen2.5-Coder 1.5B and Qwen2.5-Coder 3.0B achieve \textit{pass@1} scores of 54\% and 59\%, respectively, with VRAM usage below 10 GB and inference times under 0.75 seconds. In contrast, OpenCodeInterpreter 6.7B and Qwen2.5-Coder 7B reach higher \textit{pass@1} scores (62\% and 65\%), but require substantially more resources. Qwen2.5-Coder 7B, for instance, achieves a 20.37\% improvement in \textit{pass@1} score over Qwen2.5-Coder 1.5B (65\% vs. 54\%), but this gain comes at the cost of a 261\% increase in VRAM usage(23.7 GB vs. 6.55 GB). On the other hand, SLMs like Incoder 6.7B and PolyCoder 2.7B are positioned toward the top-left with a larger bubble, indicating lower accuracy, slower inference, and higher resource requirements.

Beyond these broader trends, several additional patterns emerge when examining the results more closely. Some models exhibit poor efficiency relative to their size, for example, PolyCoder 2.7B requires 14.15 GB of VRAM and has the slowest inference time in the entire study (5.58 seconds), yet achieves only 0.03 \textit{pass@1}. Similarly, CodeLlama 7B consumes 23.58 GB of VRAM, comparable to other Group 3 models, but delivers only 0.37 \textit{pass@1}, making it one of the least efficient models in terms of accuracy per memory cost. Conversely, several small and mid-sized models demonstrate highly favorable trade-offs. Phi-1 1.3B achieves 0.41 \textit{pass@1} with only 8.45 GB VRAM and a 1.28 second inference time, outperforming multiple larger models, while StarCoder2 3B delivers 0.42 \textit{pass@1} with just 7.99 GB VRAM and a 1.03 second inference time. These comparisons highlight that parameter size alone does not predict efficiency and that well-optimized architectures and training pipelines can allow smaller SLMs to rival or surpass larger ones in practical performance.

Further, we statistically evaluate whether inference speed and resource requirements differ meaningfully across model groups by conducting a one-way ANOVA test. Table~\ref{tab:anova_resource} reports the results of the one-way ANOVA test. The results show that VRAM usage varies significantly between groups (p \(=\) 0.000002). However, no statistically significant difference was found in inference time (p = 0.223). To further identify which group differences in VRAM usage are statistically significant, we conducted Tukey’s HSD post-hoc test, as reported in Table~\ref{tab:tukey_vram_posthoc}. The results show that the VRAM requirements of Group 3 models are significantly higher than those of Group 1 and Group 2. Specifically, the mean difference between Group 1 and Group 3 is 12.27 GB with a p-value of $<$ 0.0001, and between Group 2 and Group 3 is 10.43 GB with a p-value of 0.0001, both indicating statistically significant differences. In contrast, the mean difference between Group 1 and Group 2 is only 1.84 GB with a p-value of 0.5922, which is not statistically significant.

\renewcommand{\arraystretch}{1.5} 

\begin{table}[H]
\caption{ANOVA Test Results for Differences in VRAM Usage and Inference Time Across Model Groups}
\centering
\resizebox{\columnwidth}{!}{%
\begin{tabular}{|l|c|c|c|}
\hline
\rowcolor{lightgray}
\textbf{Metric} & \textbf{F-statistic} & \textbf{p-value} & \textbf{ Significant  ($p{<}0.05$)?} \\
\hline
Mean VRAM Usage     & 32.10 & 0.000002 & YES \\
Mean Inference Time & 1.64  & 0.223571 & NO \\
\hline
\end{tabular}
}
\label{tab:anova_resource}
\end{table}

\renewcommand{\arraystretch}{1.5} 

\begin{table}[H]
\caption{Tukey’s HSD Post-hoc Test for Pairwise Comparison of Model Groups on VRAM Usage}
\centering
\resizebox{\columnwidth}{!}{%
\begin{tabular}{|c|c|c|c|c|c|}
\hline
\rowcolor{lightgray}
\textbf{Metric} & \textbf{Candidate 1} & \textbf{Candidate 2} & \textbf{Mean Diff} & \textbf{p-adj} & \textbf{Significant} \\
\hline
\multirow{3}{*}{VRAM usage}
& Group 1 & Group 2 & 1.8391  & 0.5922 & No \\
& Group 1 & Group 3 & 12.2679 & 0.0000 & \textbf{Yes} \\
& Group 2 & Group 3 & 10.4288 & 0.0001 & \textbf{Yes} \\
\hline
\end{tabular}
}
\label{tab:tukey_vram_posthoc}
\end{table}

\begin{tcolorbox}[colback=gray!10, colframe=black, title=Key Finding from RQ2, left=0pt, right=0pt, boxsep=5pt]
\begin{itemize}
    \item Larger SLMs tend to deliver higher code generation performance, but this comes at the cost of significantly increased memory usage, while inference time remains statistically similar across model sizes. Nevertheless, some smaller and mid-sized models demonstrate strong performance with notably lower resource demands, making them attractive for memory-constrained environments.
\end{itemize}
\end{tcolorbox}

\subsection{Multilingual Consistency (RQ3)}
\label{MultilingualConsistency(RQ3)}

\begin{figure*}[h]
    \centering
    \setlength{\tabcolsep}{0pt} 
    \renewcommand{\arraystretch}{0} 
    \begin{tabular}{cc}
        \begin{minipage}[b]{0.5\textwidth}
            \includegraphics[width=\linewidth]{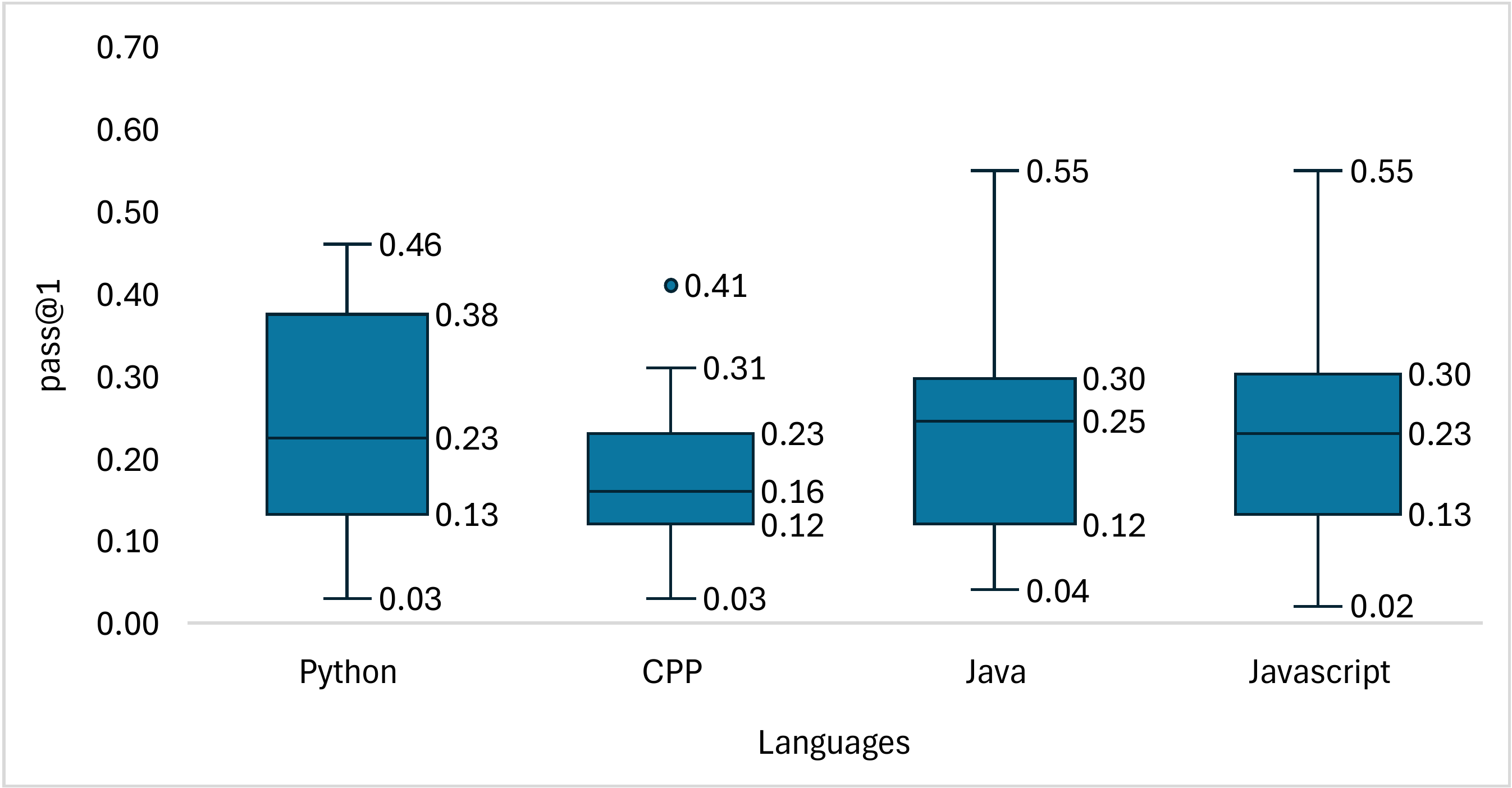}
        \end{minipage} &
        \begin{minipage}[b]{0.5\textwidth}
            \includegraphics[width=\linewidth]{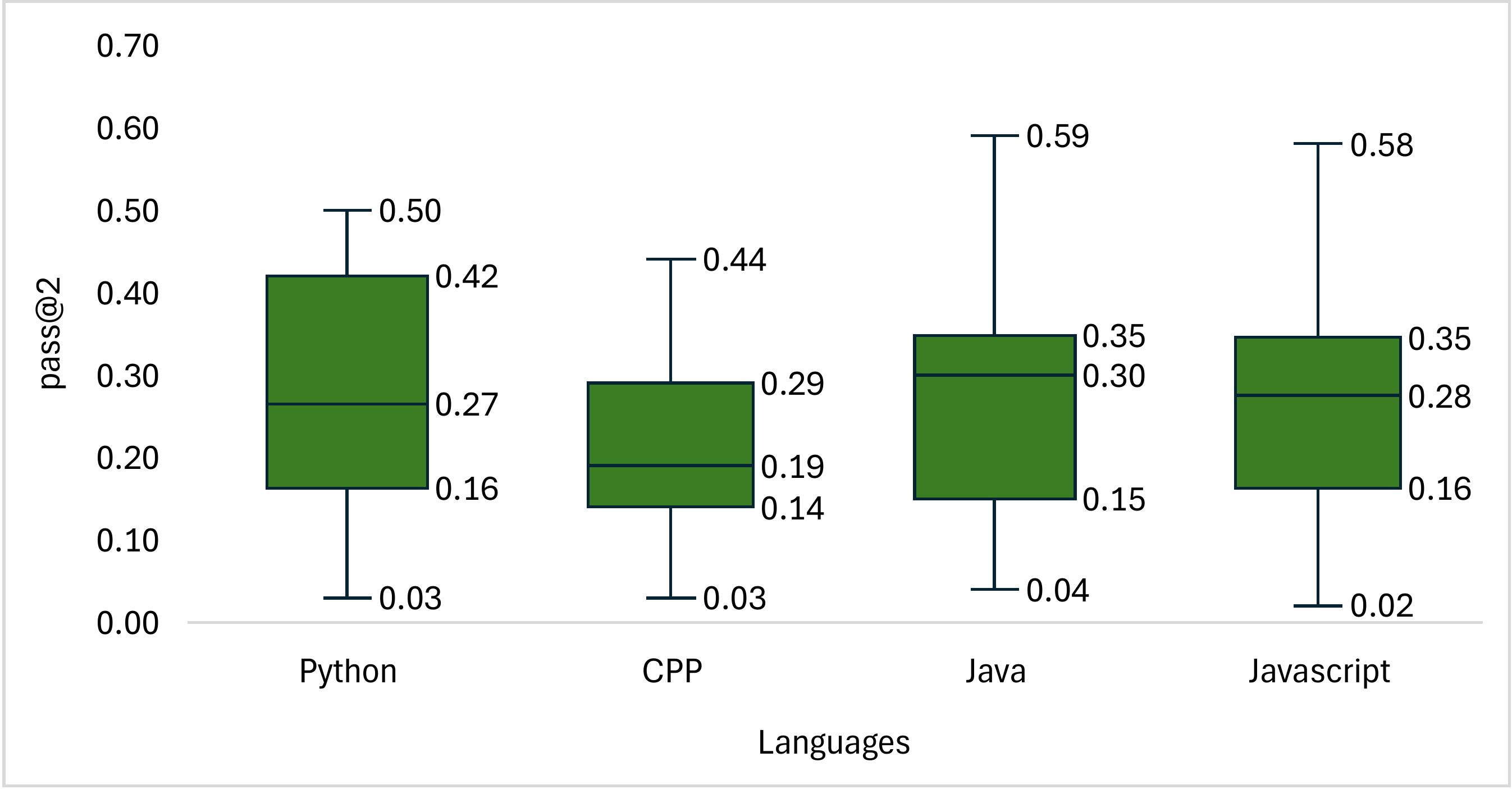}
        \end{minipage} \\
        \begin{minipage}[b]{0.5\textwidth}
            \includegraphics[width=\linewidth]{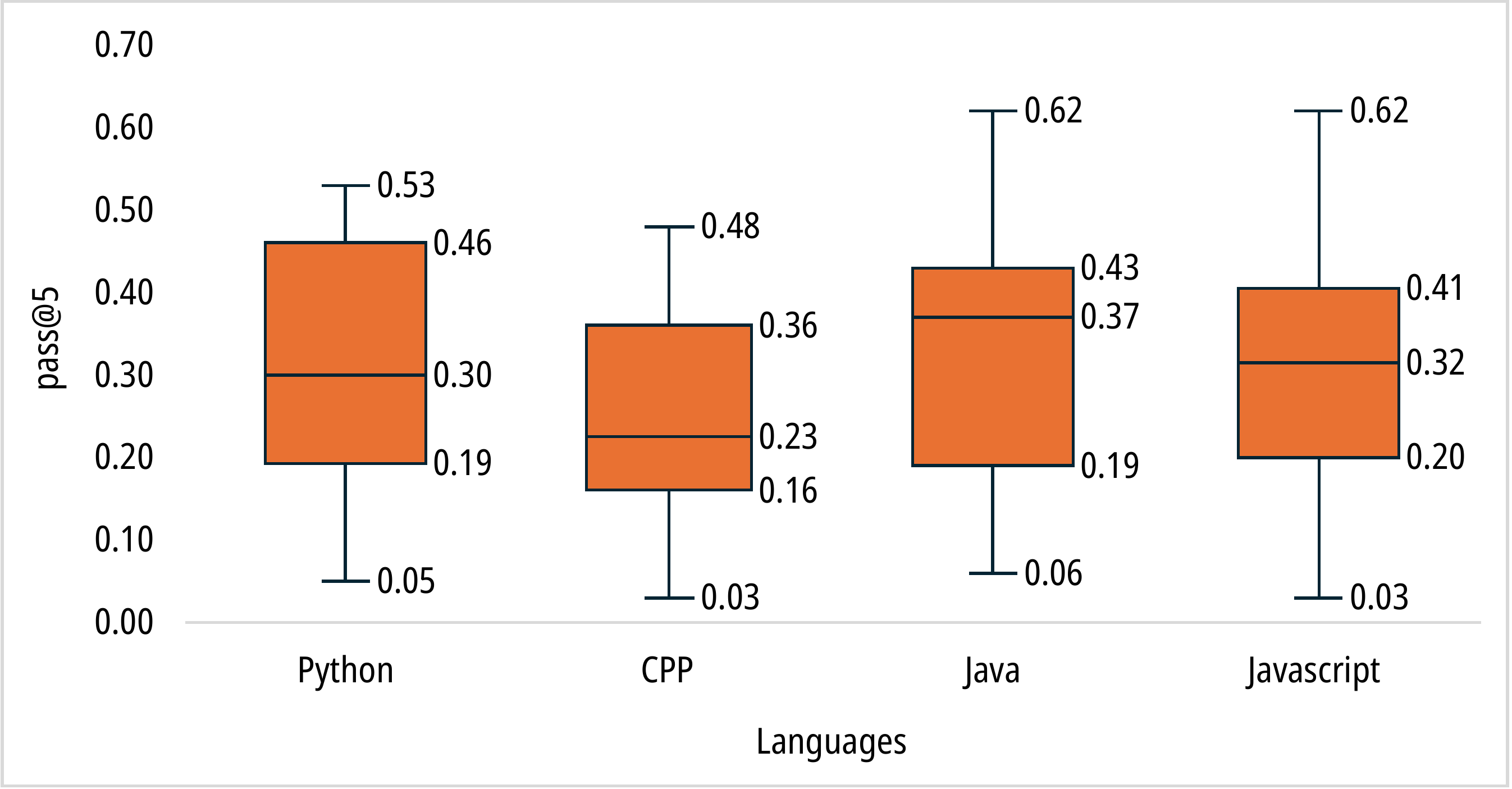}
        \end{minipage} &
        \begin{minipage}[b]{0.5\textwidth}
            \includegraphics[width=\linewidth]{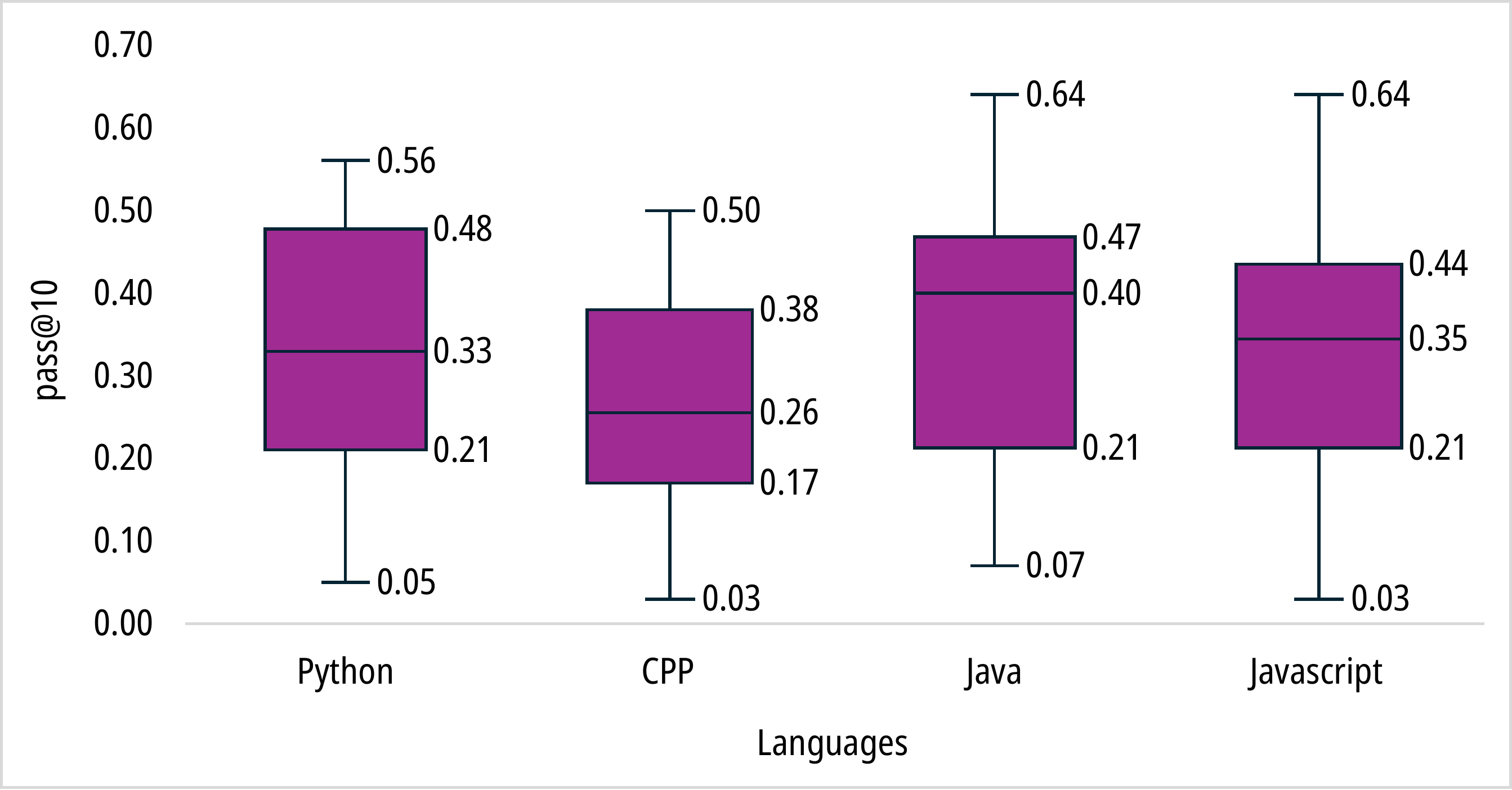}
        \end{minipage}
    \end{tabular}
    \caption{Distribution of \textit{pass@k} scores across programming languages in the HumanEvalPack benchmark.}
    \label{fig:RQ3_HEpack}
\end{figure*}

We examine the multilingual performance of the models on two multilingual benchmarks: HumanEvalPack and CodeXGLUE. These benchmarks evaluate functional correctness and lexical similarity across multiple programming languages, providing insight into how model performance varies across different languages. Figure~\ref{fig:RQ3_HEpack} presents the distribution of \textit{pass@k} scores across four languages, i.e., Python, C++, Java, and JavaScript, in HumanEvalPack. The box plots display the median, interquartile range, and outliers for each language. At \textit{pass@1}, Java achieves the highest median score (0.25), followed by Python and JavaScript (both 0.23), and C++ (0.16). This relative ranking remains consistent at higher k-values. For instance, at \textit{pass@10}, Java reaches a median of 0.40, compared to 0.33 for Python, 0.35 for JavaScript, and 0.26 for C++. C++ consistently shows the lowest performance, particularly with narrow interquartile ranges and low upper bounds, indicating limited variation and poor model accuracy for this language. This behaviour aligns with the fact that C++ problems often involve stricter type constraints and more complex syntax than other languages, which increases the difficulty of generating functionally correct code. As there are observable differences in model performance across programming languages, we conducted a one-way ANOVA test to evaluate whether these differences are statistically significant. Table \ref{tab:anova_passk} presents the results of a one-way ANOVA test conducted on the \textit{pass@k} scores obtained from the HumanEvalPack benchmark. The p-values are 0.46 for pass@1, 0.39 for pass@2, 0.32 for pass@5, and 0.27 for pass@10. These values are greater than 0.05, indicating that there is no statistically significant difference in model performance across the four programming languages evaluated.

\begin{figure*}[h]
\centering
\includegraphics[width=\linewidth]{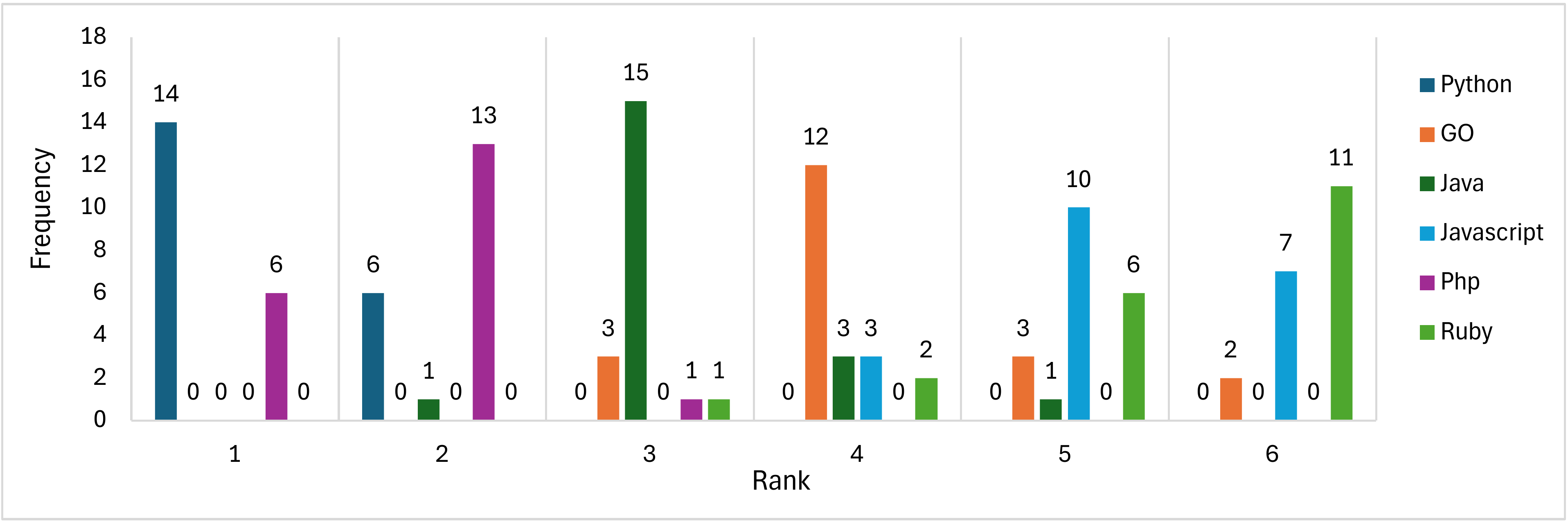}
\caption{Frequency of programming languages appearing in top BLEU score rankings in the CodeXGLUE Benchmark}
\label{fig:RQ3_codex}
\end{figure*}

In addition to \textit{pass@k} based functional correctness, we also analyse natural language generation quality using BLEU scores on the CodeXGLUE benchmark. Figure~\ref{fig:RQ3_codex} shows how frequently each language appears among the top in BLEU score ranks across the 20 models. Python appears in Rank 1 for 14 models and Rank 2 for 6 models, making it the most dominant language in high BLEU score positions. PHP ranks first in 6 models and second in 13, indicating strong performance among models in lower-ranked positions as well. Java dominates Rank 3, appearing 15 times, while Go appears most frequently in Rank 4 with 12 instances. Ruby appears 11 times in Rank 6 and 6 times in Rank 5, showing a strong presence in lower ranks. In contrast, JavaScript appears only 3 times in Rank 4, 10 times in Rank 5 and 7 times in Rank 6. It does not feature at all in the top two positions. These results highlight a clear skew toward Python and PHP in BLEU score rankings, while Java, Go, Ruby, and JavaScript frequently occupy mid-to-lower ranks. Python and PHP achieve higher BLEU ranks because their documentation styles in CodeXGLUE are concise, making them easier for SLMs to match lexically. In contrast, languages like Java, Go, Ruby, and JavaScript use more variable documentation patterns, which naturally lowers lexical overlap. Since BLEU measures surface similarity rather than semantic quality, these differences reflect documentation style and training-data distribution rather than inherent model weakness in those languages.

\renewcommand{\arraystretch}{1.3} 

\begin{table}[!H]
\caption{ANOVA Test Results for \textit{pass@k} Scores Across Programming Languages in HumanEvalPack benchmark}
\centering
\begin{tabular}{|l|c|c|c|}
\hline
\rowcolor{lightgray}
\textbf{Metric} & \textbf{F-statistic} & \textbf{p-value} & \textbf{Significant ($p{<}0.05$)?} \\
\hline
pass@1  & 0.88 & 0.46 & NO \\
pass@2  & 1.02 & 0.39 & NO \\
pass@5  & 1.20 & 0.32 & NO \\
pass@10 & 1.34 & 0.27 & NO \\
\hline
\end{tabular}
\label{tab:anova_passk}
\end{table}

\begin{tcolorbox}[colback=gray!10, colframe=black, title=Key Finding from RQ3,  left=0pt, right=0pt, boxsep=5pt]
\begin{itemize}
  \item Models generally perform better in languages such as Python, Java, and PHP, while showing weaker performance in languages like Go, C++ and Ruby. However, statistical analysis indicates that these differences are not significant.
\end{itemize}
\end{tcolorbox}

\section{Discussion}
\label{Discussion}

This section discusses key findings for each research question and outlines their implications for research and practice. It highlights how model size affects performance and stability, the trade-offs between efficiency and accuracy, and the challenges of achieving consistent multilingual performance in code generation.

\subsection{Model Size on Code Generation Quality (RQ1)}
\textbf{Effect of Model Size and Benchmarks on Code Generation Performance}: Our results demonstrate that model size significantly impacts performance, whereas benchmark choice does not lead to statistically significant differences. This suggests that the relative strengths and weaknesses of SLMs remain consistent across different code generation tasks, such as those represented by HumanEval, MBPP, and Mercury. It implies that the models generalisation ability is largely governed by their architectural and capacity-related factors rather than task-specific peculiarities.

This observation aligns with prior empirical studies. For example, Codex by Chen et al.\cite{Chen2021EvaluatingLL} shows a strong correlation between model size and \textit{pass@k} accuracy in code generation. Peng et al.\cite{peng2024humaneval} similarly observed minimal variation in benchmark performance compared to variation due to model size, suggesting that parameter scaling is a dominant driver of effectiveness. In addition, Ye et al.\cite{ye2024benchmarking} emphasised that generalisation across code tasks becomes robust once a model crosses a certain size threshold, with benchmark differences becoming negligible. These prior findings highlight a reliable cross-task behaviour of SLMs, especially when evaluated at scale. The lack of benchmark sensitivity suggests that improvements in model design and training translate well across tasks with differing structural and semantic requirements. It further implies that benchmark-agnostic development and evaluation strategies may be valid for SLMs, particularly when considering architectural scaling or transfer learning.

\textit{\textbf{Implications}}: Our findings offer researchers an opportunity to focus on architecture design, scaling laws, and general-purpose training strategies that extend model generalisation beyond individual benchmarks. For practitioners, the benchmark-agnostic nature of performance simplifies deployment decisions, allowing the use of a single, well-performing SLM across various programming tasks without the need for fine-tuning or adaptation.

\textbf{Performance and Stability of Large vs. Smaller SLMs}: The obtained results of our evaluation show that larger SLMs tend to outperform smaller SLMs in both functional correctness and stability. They generally occupy top positions across evaluation metrics, indicating strong performance and consistent behaviour across diverse code generation tasks. However, our data also reveal that this trend does not always hold true; for example, some small and mid-sized models (e.g., OpenCodeInterpreter 1.3B, Qwen2.5-Coder 3B) challenge it. Some models from lower parameter groups consistently secure high rankings and demonstrate notable performance and stability. These findings suggest that while scaling up the number of parameters generally improves models' performance, it is not the only way to achieve strong performance in code generation. Well-designed and trained smaller SLMs can achieve competitive results when optimised effectively.
\newline
Similar patterns have been documented in earlier empirical work. For example, Ye et al.\cite{ye2024benchmarking} observed that once SLMs surpass a certain parameter threshold, additional gains in functional correctness become marginal. This resonates with our finding that while larger models generally perform well, several smaller and mid-sized models also achieve high accuracy and stability, suggesting diminishing returns from parameter scaling alone. It indicates that at some point, it becomes more beneficial to focus on how the model is trained or what data it utilises, rather than merely increasing its size. Similarly, Xu et al.\cite{xu2024survey} and Wang et al.\cite{Wang2024ACS} emphasised the feasibility of achieving competitive performance from well-optimised SLMs, especially in resource-constrained environments. Our results support this by demonstrating that small models, such as Qwen2.5-Coder 3.0, can outperform several larger models. Furthermore, Subramanian et al.\cite{Subramanian2025survey} emphasised the underappreciated potential of SLMs in real-world Software Engineering workflows, calling for more empirical performance assessments that go beyond size alone. Our evaluation aligns with this perspective by showing that model size alone is not a reliable predictor of practical utility or performance consistency, especially in diverse code generation scenarios. 

Although this study focuses on SLMs in the 0.4B–10B parameter range, it is useful to briefly situate these findings with LLMs. Previous studies by Zheng et al. \cite{zheng2024well} report results for GPT-3.5 on the HumanEval benchmark using settings such as temperature = 0.4 and a single sample (n = 1) per problem. On the other hand, Li and Murr \cite{xia2024top}evaluate recent GPT models on HumanEval in a zero-shot setup for n=10 samples via the OpenAI API, and Chen et al. \cite{Chen2021EvaluatingLL} report HumanEval results for Codex with sample generation at temperature = 0.8 and up to n=100 samples per problem. However, these studies differ from ours in several important aspects, including the experimental pipeline and configurations used, the set of LLMs evaluated, and the benchmark coverage. Prior work typically reports results for few models on one or two benchmarks (e.g., HumanEval or MBPP), whereas our study evaluates 20 SLMs across five distinct benchmarks. Due to these methodological differences, a direct quantitative comparison would not be reliable. Nevertheless, at a high level, comparing our results with existing results suggest that well-designed and carefully trained SLMs can achieve performance comparable to LLMs under specific benchmark conditions. This observation indicates that higher parameter count alone is not a dependable indicator of code generation quality, and that architectural choices, training procedures, and data curation play a critical role in determining practical performance.

\textbf{\textit{Implications}:} Our findings highlight that high code generation performance is not solely dependent on parameter count. For researchers, this opens up opportunities to explore alternative strategies such as architectural refinements, training procedure optimisations, and task-specific data curation. The ability of smaller and mid-sized SLMs to achieve competitive performance suggests that research efforts can focus on enhancing model efficiency and specialisation without relying on large-scale scaling. For practitioners, the strong performance of smaller models provides a compelling case for adopting SLMs in settings with constrained resources. This is particularly relevant for deployment on edge devices or in environments with limited capacity. The presence of high-performing models in smaller sizes reinforces the viability of cost-effective and resource-efficient model selection strategies, enabling practical deployment of SLMs in real-world software engineering environments without compromising performance.

\subsection{Efficiency Trade-offs in SLM Deployment (RQ2)}
\textbf{Resource Consumption and Inference Speed vs. Performance of SLMs:} Our analysis reveals that GPU memory usage (VRAM) increases significantly with model size. At the same time, inference speed does not show a corresponding statistical variation. We observe this behavior because all models in our study are SLMs (0.4B–10B parameters), where differences in latency are naturally smaller than in large LLMs (e.g., >30B parameters). In addition, all models were evaluated using identical decoding settings and on relatively small, single-function benchmark tasks, which further reduces variation in inference speed.  This indicates that, rather than inference speed, VRAM is the constraint when deploying SLMs in practice. As models scale up, their memory demands grow higher, creating significant challenges for deployment in resource-constrained environments. However, this does not imply that larger models are always necessary to achieve high performance. While high-performing SLMs often belong to the larger parameter groups, several smaller and mid-sized models demonstrate a favourable balance between correctness and resource efficiency. These models offer compelling alternatives for use cases where hardware limitations prohibit the use of resource-intensive models. 
\newline
Prior studies (e.g.,\cite{Lu2024SmallLM},\cite{schick-schutze-2021-just}) have similarly emphasized this dynamic. While inference speed can often be optimised through software and hardware tuning, memory capacity, especially VRAM, remains a primary constraint in deploying small language models.  Recent findings by Ashkboos et al. \cite{ashkboos2024computational} indicate that even SLMs with modest parameter counts face significant memory bottlenecks during both training and inference, limiting their usability in resource-constrained settings. Lu et al.\cite{Lu2024SmallLM} also observed that many performance evaluations overlook memory bottlenecks, leading to unrealistic assessments of deployment feasibility. Moreover, Subramanian et al.\cite{Subramanian2025survey} emphasised treating efficiency not merely as another dimension of scale, but as an important metric in empirical evaluations of SLMs, especially in applied software engineering contexts. Furthermore, Schick and Schütze \cite{schick-schutze-2021-just} suggest that small models can be efficient few-shot learners, particularly when designed for narrow domains or specialised tasks. Observations from these studies are aligned with our findings, which demonstrate that several well-optimised small and mid-sized models achieve strong correctness while maintaining low memory usage. This supports the idea that well-targeted and task-adapted SLMs can offer practical performance advantages without the high memory demands typically associated with larger models. 

\textbf{\textit{Implications}:} These findings suggest that researchers should focus more on designing models that use memory efficiently, especially when working with small or mid-sized SLMs. Instead of only trying to improve performance by increasing model size, it is important to explore new ways to make smaller models smarter through better training methods, optimised architectures, or carefully selected data for training. By doing so, researchers can help build models that perform well while also being easier to run on standard hardware. For practitioners, this means there are good options available even when resources are limited. Smaller and mid-sized SLMs that perform well while using less memory make it possible to utilise AI tools in a wider range of settings, such as edge devices like mobile phones or low-cost development environments. This supports the more practical and cost-effective adoption of SLMs in real-world software engineering, without relying on large, resource-intensive models.

\subsection{Multilingual Capabilities and Limitations of SLMs (RQ3)}
\textbf{Performance Analysis of SLMs in Multilingual Code Generation:} Our multilingual performance analysis shows that the performance of SLMs appears to fluctuate across programming languages; however, the analysis reveals that these variations are not statistically significant. This suggests that overall, SLMs maintain a reasonably consistent ability to generalise across multiple programming languages. However, the multilingual benchmarks used in this study cover only a limited set of widely used programming languages; as a result,  this generalizability applies only to the languages included in our study and cannot be assumed to extend to all programming languages. Nevertheless, practical patterns are still observable as models tend to perform better in languages such as Python, Java, and PHP, while showing comparatively weaker results in Go, C++ and Ruby. The possible reasons for these trends include: (i) the uneven representation of programming languages in pretraining datasets, and (ii) variations in syntactic structure across languages. 

Prior studies have shown that languages like C++ and Ruby introduce greater syntactic depth, semantic ambiguity, and structural irregularity, which pose challenges for code generation models to handle consistently. Raychev et al. \cite{Raychev} explained that C++ is a complex language because of features like templates and unclear behaviour in some situations, which can make it harder for models to generate correct code. Similarly, Allamanis et al. \cite{Allamanis} noted that unusual grammar rules and language-specific features in languages like Ruby can make it more challenging for models to perform well when working with multiple programming languages. Wang et al. \cite{wang2021codet5} also observed that performance on different languages in code generation tasks can be strongly influenced by how well those languages are represented in the training data and by the model's ability to generalize across structurally diverse programming paradigms. 

\textbf{\textit{Implications}:} For researchers, the lack of statistical significance should not be interpreted as complete equality across languages. Instead, it highlights the need for more targeted studies into how structural and linguistic properties of different programming languages interact with model architecture and pretraining data. Such studies can help uncover hidden biases or limitations and inform improvements in tokeniser design, data sampling, or multilingual training strategies. For practitioners, our results suggest that SLMs can be generally relied upon to work across most common languages; however, extra care may be needed when deploying models in environments dominated by more complex or underrepresented languages, such as C++ or Ruby. Additional validation, domain-specific fine-tuning, or fallback mechanisms may help ensure consistent reliability in those contexts.

\section{Threats to Validity}
\label{ThreatstoValidity}
We recognise several potential threats to the validity of this study, which we discuss below in terms of construct, internal, external, and conclusion validity.

\subsection{Construct Validity}

Construct validity addresses whether our evaluation accurately captures the intended outcomes; in this study, the code-generation capabilities of SLMs. Our evaluation uses three established benchmarks: HumanEval, MBPP, and Mercury. These benchmarks provide a controlled, reproducible method for assessing functional correctness using the \textit{pass@k} metric. However, the tasks in these benchmarks are generally small-scale, algorithmic in nature, and primarily focused on Python. This means they represent only a limited portion of real-world software engineering, which commonly requires multi-module codebases, asynchronous workflows, domain-specific APIs, and interaction with legacy systems. This limitation narrows the scope of our evaluation. To partially mitigate this constraint, we complement these benchmarks with multilingual benchmark datasets to reduce language-specific bias in the evaluation, although this does not fully eliminate the underlying constraint.


To expand our study coverage beyond Python-only evaluation, we included HumanEvalPack and CodeXGLUE benchmarks for multilingual assessment. This strategy broadens our analysis to several programming languages. However, both benchmark suites are still dominated by high-resource languages such as Python, Java, JavaScript, PHP, C++, and Go. Languages with smaller or rapidly evolving ecosystems, such as Rust, Kotlin, Swift, and TypeScript, are largely absent. As a result, the multilingual generalization observed in our results applies only to the languages included in these datasets. To address this limitation, we interpret our multilingual findings within the scope of the supported languages and avoid extending claims to broader cross-language generalization.


Furthermore, the use of automated metrics defined by the benchmark suites enables reproducible comparisons across a large number of models. However, it influences construct validity. \textit{Pass@k} measures functional correctness, while BLEU captures surface-level similarity rather than deeper semantic or stylistic qualities. These metrics do not account for aspects of code quality such as maintainability, readability, performance efficiency, idiomatic usage, or security. Additionally, our study does not include human judgments of code quality, qualitative assessments of readability or usability, or triangulation with real-world software development data such as pull requests, bug reports, or production code artifacts. We acknowledge that the current scope of our study may not capture several dimensions of code quality that are important to practitioners. This limitation points to an opportunity for future work to integrate richer evaluation methods, such as human code reviews, static analysis tools, security assessments, and project-level tasks to achieve a more holistic understanding of SLM capabilities.

An additional concern for construct validity arises from potential contamination by benchmarks. Models can have been trained on datasets that include tasks similar or identical to the benchmark used in this study. To address this concern, we examined the models' training data. Although some model families (e.g., Qwen2.5-coder) explicitly report filtering out datasets such as HumanEval and MBPP during pretraining, the full pretraining data of many models are not well documented. As a result, we cannot rule out this risk. Such overlap could inflate performance for certain models. To mitigate this limitation, we follow standard benchmarking practices used in prior work and acknowledge that potential training data leakage remains a limitation of our evaluation.

Finally, our use of static benchmarks and fully automated evaluation procedures provides consistency and reproducibility but may not reflect dynamic or interactive development workflows. Modern software engineering often involves iterative refinement, debugging, code review, and adaptation to evolving codebases—capabilities that static benchmark tasks cannot fully capture. Although these design choices enabled rigorous comparisons across models, they inevitably constrain the construct space represented in this study. While our benchmark suite is well-accepted and diverse, we acknowledge that it does not encompass the full range of capabilities required for robust, production-grade code generation. This limitation cannot be fully addressed within the scope of this work, and future research should incorporate more interactive and context-rich evaluation settings to better approximate real-world development scenarios.

\subsection{Internal Validity}
Internal validity for this study refers to the accuracy with which we conducted our experiments and to the extent to which we can trust the resulting data. To ensure fair and consistent evaluation, we carefully designed our experiments with fixed decoding settings (e.g., temperature = 0.2, top-p = 0.95, max tokens = 512), a standard zero-shot prompting format, and automated evaluation tools. These decoding parameters follow standard practice in code-generation benchmarks and match widely used configurations (e.g., official HumanEval implementations and MBPP evaluation scripts \cite{openai_humaneval_2021}, \cite{google_mbpp_2021}), as well as vendor recommendations that suggest low-temperature sampling for deterministic code generation (e.g., NVIDIA NeMo documentation \cite{nvidia_nemo_2025}). 
A limitation of this approach is that decoding parameters can affect generation diversity and correctness, leading to performance variations across configurations. Although a full sensitivity analysis could provide a more detailed understanding of how decoding settings affect model behavior, conducting such an analysis across all models would require exploring a large multidimensional parameter space, substantially increasing computational cost and reducing comparability with prior benchmark studies. To maintain methodological consistency, we therefore applied the same decoding configuration across all models. We acknowledge this as a limitation of this study, and future research should systematically examine the robustness of SLM performance under different decoding parameters.

We used a zero-shot prompting strategy for all tasks, meaning the models received no in-context examples. Although this setup reflects a generalization, it may underestimate model performance relative to few-shot prompting or instruction tuning, which are commonly used in practice. We choose this strategy because prior work shows that few-shot prompting predominantly benefits instruction-tuned models rather than base models. As all models in our study are based on SLMs, incorporating few-shot examples would introduce methodological bias and reduce comparability. Nonetheless, we acknowledge that an evaluation relying exclusively on zero-shot prompting limits the ability to assess model performance in example-driven or interactive development workflows. As a result, this remains a limitation of our evaluation: future work should evaluate SLMs across different prompting settings. Besides, our experiments were conducted on two different hardware platforms: an NVIDIA L4 GPU and an RTX 3090 GPU, each equipped with 24 GB of VRAM. This ensures that the primary efficiency metric (VRAM usage) in our analysis remains hardware independent. Although similar in memory, architectural differences such as CUDA cores, memory bandwidth, or driver optimisations can influence inference latency. To mitigate this issue, each model was evaluated entirely on a single GPU type, and models were evenly distributed across both platforms. In addition, both hardware environments used identical software stacks (CUDA version, driver version, batch sizes, decoding settings), and we fixed random seeds to reduce variability. These precautions ensure that any hardware-related variability appears as random noise rather than systematic bias. Besides, we acknowledge that using a single GPU type would be ideal. Still, the steps taken help ensure that hardware differences do not meaningfully affect the reliability of our comparative results.



\subsection{External Validity}
External validity refers to the extent to which our results generalise beyond the specific experimental setup. In our study, we only included models with a decoder-only architecture. All of them were open-source and had up to 10 billion parameters. They were released between 2022 and 2024. This means we did not include other model types, such as encoder-only or encoder-decoder models. We also did not evaluate larger models or proprietary ones, such as GPT-4 or Gemini. As a result, our findings are limited in scope and may not apply to other types of models. This design choice was intended to ensure comparability across open-source SLMs under similar resource constraints, but it limits generalisation to other model classes. 

Another threat to external validity arises from the nature of the tasks used in this study. The selected benchmarks consist primarily of short, self-contained programming problems that do not capture key characteristics of real-world software engineering, such as multi-file project structures, long-range dependencies, integration with external libraries, or iterative development workflows. While these benchmarks are widely used in code-generation research and provide a controlled, reproducible evaluation environment, the evaluation results represent only a partial view of practical software development. We address this limitation in part by including multiple benchmark spanning different task types and languages; however, static benchmarks cannot fully reflect the complexities of production environments, where broader contextual understanding and architectural reasoning are required. We therefore acknowledge this as a limitation of the present work, and future research should incorporate project-level datasets or tasks drawn from real-world repositories to assess SLM behavior in practical settings more comprehensively. Moreover, the benchmarks used in this study evaluate SLMs on only a subset of languages and exclude niche or domain-specific languages common in specialized software engineering contexts. We partially address this limitation by incorporating multilingual benchmark suites, such as HumanEvalPack and CodeXGLUE, which extend coverage beyond Python and enable evaluation across multiple widely used languages. However, these benchmarks still represent only a limited portion of the broader programming language ecosystem. As a result, our multilingual findings should be interpreted within the scope of the languages included in these datasets and may not generalize to specialized or domain-specific languages. We acknowledge this limitation, and future work should evaluate SLMs on more diverse, domain-relevant language sets to improve generalizability

A further limitation concerns the efficiency dimensions evaluated in this study. Our assessment of computational efficiency focused on VRAM usage and inference time, as these metrics are commonly reported in prior work and can be measured reliably in GPU-based environments. However, this approach does not capture other important aspects of efficiency, such as energy consumption, CPU utilization, thermal characteristics, or power efficiency. These factors are particularly relevant for edge devices and low-power deployment scenarios \cite{s23031279}. These dimensions were outside the scope of this work. As a result, our findings represent only a partial view of the practical deployment considerations for SLMs. We acknowledge this limitation and identify the inclusion of energy profiling and hardware-level performance measurements as an important direction for future research to provide a more comprehensive understanding of real-world deployment.

In addition, we used zero-shot prompting across all tasks to test how well models generalise without seeing any examples. However, it may not reflect many real-world coding scenarios, where developers often provide examples, context, or iterative feedback. Consequently, models optimised for instruction-following or few-shot learning may appear weaker under zero-shot evaluation. As a result, our evaluation may not accurately reflect how well some models perform when provided with additional guidance. To mitigate this, we used a standardised zero-shot format across all models to ensure fair comparisons, but future work should include few-shot and interactive prompting setups to approximate practical usage better. Models that rely heavily on context may appear weaker in this setup. While zero-shot evaluation helps reveal a models raw ability, it may limit our understanding of its full potential in interactive settings.



\subsection{Conclusion Validity}
Conclusion validity describes how well our results support the drawn conclusions. In this study, we employed statistical methods, including ANOVA and Tukey's HSD post hoc test, to assess whether differences in performance and efficiency across models and benchmarks were statistically significant. These tests come with certain assumptions. First, the data points should be independent, which was satisfied since each model was evaluated separately. Second, the results within each group should exhibit similar variance, which we achieved by grouping models by size and maintaining consistent conditions. Third, the data should follow a roughly normal distribution, which is reasonable in our case because we used averages across many tasks. Lastly, the model groups were not randomly selected but carefully chosen, which meets the fixed-effects requirement of ANOVA. Given how we set up the experiments, all these assumptions were reasonably met. However, several factors may still affect the results, such as timeouts or out-of-memory errors, particularly when larger models are tested on tasks that require more memory. Such failures may result in incomplete results for some models on specific benchmarks, potentially introducing bias. To reduce this issue, we adjusted batch sizes and limited the number of token generations where needed. We also documented any failed runs and excluded them consistently across models to avoid skewing comparisons. Some efficiency metrics, such as inference speed, may also be affected by slight differences in system load or hardware behavior. To mitigate this, we ensured that each model was run under controlled and consistent system configurations, with dedicated runs to minimize interference from background processes. Although some slight variations may still exist, we observed consistent trends, suggesting that our conclusions are reliable and would likely hold under similar conditions.

\section{Related Work}
\label{RelatedWork}
With the rapid advancements in Artificial Intelligence (AI), Language Models have emerged as powerful tools for code generation \cite{zhang2023survey}. Although LLMs such as GPT-4 \cite{GPT-4} and Gemini \cite{team2023gemini} have demonstrated impressive capabilities, they continue to face practical challenges. Xu \textit{et al.}\cite{xu2024survey} conducted a comprehensive survey on resource-efficient LLMs and highlighted that both training and inference require high-performance GPUs, substantial memory bandwidth, significant power, and specialised cooling systems, which limit their widespread adoption. Wang \textit{et al.} \cite{Wang2024ACS} discussed the use of SLMs as a more practical alternative. They showed that SLMs can perform well on specific tasks while using less memory. The authors also explained that SLMs are easier to deploy and can work better in real-world settings where computing resources are limited. However, our current understanding of the consistency and effectiveness of SLMs in the context of code generation remains underexplored.

Recently, many LLMs have been explicitly developed focusing on coding and have been widely applied to tasks such as program synthesis, code completion, and bug fixing \cite{chen2024survey}. Chen \textit{et al.}\cite{Chen2021EvaluatingLL} introduced Codex, a code-specific LLM trained in extensive public code repositories. They discussed its ability to convert natural language instructions into executable code and reported its strong performance on the HumanEval benchmark, highlighting its practical utility in tools such as GitHub Copilot. Rozière \textit{et al.} \cite{CodeLlama} introduced an open source alternative, Code Llama, offering competitive results in benchmarks and support for extended input contexts. The authors reported that its performance on the HumanEval and MBPP benchmarks is on par with some of the strongest open models available, particularly in zero-shot settings. Furthermore, Li \textit{et al.} \cite{Li@2022} introduced AlphaCode, which was tested on competitive programming challenges. They emphasised its ability to produce diverse and creative solutions that aligned closely with human-level problem-solving strategies, achieving results comparable to those of skilled human participants.

In response to growing concerns about the resource demands of LLMs, recent studies have focused on developing SLMs that are more efficient and accessible. Wang \textit{et al.} \cite{Wang2024ACS} emphasized that SLMs can significantly reduce the computational overhead while still achieving competent performance on various tasks. Among early SLMs, GPT-Neo \cite{gpt-neo} was introduced as a lightweight alternative capable of general-purpose code generation, with the authors highlighting its ability to run on limited hardware while maintaining competitive performance. Similarly, Xu \textit{et al.} \cite{xu2022systematic} conducted a systematic evaluation of code-focused LMs and noted that smaller models like PolyCoder were able to perform well in certain code understanding and generation scenarios, despite having significantly fewer parameters than traditional LLMs. More recently, Lozhkov \textit{et al.} \cite{lozhkov2024starcoder} introduced StarCoder2 as a scalable open-source family of code models designed to balance performance and efficiency, and reported that its smaller variants demonstrated promising results in zero-shot code generation tasks. 

A variety of benchmarks have been developed to evaluate the performance of language models in code generation, each targeting different aspects of functionality and generalisation. Hendrycks \textit{et al.} \cite{hendrycks2021measuring} introduced APPS, a dataset designed to assess problem-solving ability through programming challenges of varying difficulty. Chen \textit{et al.} \cite{Chen2021EvaluatingLL} proposed HumanEval, a widely used benchmark where models are tested on their ability to generate correct Python functions from descriptions written in natural language. To further assess real-world programming capabilities, Austin \textit{et al.} \cite{austin2021program} released MBPP, which focuses on Python code snippets and includes manually-written test cases to evaluate correctness. These datasets have become standard tools for assessing LLMs and SLMs alike. Muennighoff \textit{et al.} \cite{OctoPack} introduced HumanEvalPack, extending these benchmarks into a multilingual setting. The authors highlighted that models trained primarily on English often struggle with syntax and semantics in different programming languages, making multilingual benchmarks essential for more comprehensive evaluations. Du \textit{et al}. \cite{Du2024MercuryAC} introduced Mercury, a benchmark specifically designed to measure code efficiency, including runtime and memory usage, in addition to correctness. The authors emphasised that efficiency remains underexplored in model evaluation, despite being crucial for practical deployment. Across these benchmarks, the most common evaluation metric is \textit{pass@k}, which measures whether at least one of the top-k generated outputs passes the unit tests. The BLEU score is also occasionally used to measure surface-level similarity between the generated and reference outputs. However, several studies caution that it may not fully capture functional correctness \cite{chen2024survey}. Chen \textit{et al.} \cite{chen2024survey} further noted that differences in evaluation protocols, such as varying sampling temperatures, prompt formats, or test coverage, make it difficult to directly compare model performance across studies. This highlights the ongoing need for standardised evaluation methods to enable more reliable and reproducible assessments of code generation models.

While much progress has been made in developing and evaluating LLMs for programming tasks, empirical studies that assess their integration and effectiveness in real-world SE workflows are still few. Zhang \textit{et al.} \cite{zhang2023survey} conducted a comprehensive survey on the use of LLMs in SE, covering areas such as program synthesis, code summarisation, and defect prediction. The authors noted that while LLMs show great promise, their actual deployment in SE practice is limited by scalability, interpretability, and cost effectiveness. Despite these efforts, most empirical studies have focused on LLMs such as Codex, with limited attention to SLMs. Subramanian \textit{et al.} \cite{Subramanian2025survey} explicitly identified this gap, noting that while SLMs are increasingly being proposed as cost-effective alternatives, their effectiveness in software engineering contexts remains largely untested. Likewise, Lu \textit{et al.} \cite{Lu2024SmallLM} observed that empirical validation of SLMs across various programming tasks is often lacking, particularly in controlled environments that reflect real-world SE needs. This points out a critical research gap: although SLMs can offer promising advantages in terms of resource efficiency and deployment feasibility, their practical applicability in SE remains underexplored and under-evaluated.

\textbf{Conclusive summary:} In summary, while LLMs have been extensively studied for code generation, there is a lack of empirical evaluation of SLMs in this context. Prior works have emphasised the potential of SLMs \cite{Wang2024ACS, Subramanian2025survey}, but their consistency and robustness in code generation tasks remain underexplored \cite{Lu2024SmallLM}. This work addresses this gap by benchmarking a range of SLMs across widely used datasets and evaluation metrics. Unlike earlier studies that focus on general NLP tasks (e.g., \cite{Lepagnol2024SmallLM}), we focus specifically on SLMs that perform code generation tasks. Our empirical study reveals their actual capabilities and limitations in resource-constrained real-world development environments. This approach enables a more informed understanding of where and how SLMs can be effectively applied in software engineering workflows.
\section{Conclusion}
\label{Conclusion}
We systematically selected and evaluated 20 open-source SLMs (see Table \ref{tab:slm-overview})  to assess their capabilities in code generation tasks. This study was designed with three RQ's defined in Section \ref{Introduction}. To address these RQ's, we conducted experiments using five widely adopted benchmarks (i.e., HumanEval, MBPP, Mercury, HumanEvalPack, CodeXGLUE) under a unified zero-shot prompting setting, with consistent decoding configurations and hardware environments. Our evaluation focused on three key dimensions. First, we assessed code generation performance by measuring how often models produced functionally correct outputs using the \textit{pass@k} metric. Second, we evaluated computational efficiency in terms of GPU memory usage and inference time to understand the resource demands of each model. Third, we examined multilingual behaviour by analysing performance across different programming languages using both \textit{pass@k} and BLEU scores. Based on the analysis of the obtained results, the key findings of this study are as follows:

\begin{itemize}
  \item 
  Our results show that strong performance in code generation is not limited to the largest SLMs. Several smaller and mid-sized models (Groups 1 and Group 2, with parameter sizes $\leq$3B) performed competitively, in some cases matching or exceeding the performance of larger SLMs (Group 3, with parameter sizes $>$3B - $\leq$10B) within our evaluation range. This suggests that thoughtful model design and training can achieve substantial performance gain without necessarily scaling up the models size to higher parameters.

  \item
  We also observed that larger models tend to require much more memory to maintain their accuracy, even though their response time is not significantly faster. This makes them harder to use in real-world environments where computing resources are limited. On the other hand, certain compact models managed to deliver both good accuracy and efficient resource use, making them attractive choices for practical deployment. 

  \item 

  Our results show that although models vary in performance across programming languages, these differences are not statistically significant. This indicates reasonably consistent behaviour within the languages included in our benchmarks. However, this should not be interpreted as broad multilingual generalisation, because our evaluation covers only a limited set of languages, and the findings apply only to those languages we tested.
  
\end{itemize}

The findings of this study offer valuable insights for both researchers and practitioners. For researchers, we discussed the opportunities to explore improved training strategies, architectural innovations, and optimisation techniques for SLMs. For practitioners, we discussed the observed trade-offs between performance and computational efficiency. It can inform practical model selection based on resource availability and task demands. Together, these insights support more effective and context-aware use of SLMs in real-world code generation scenarios. As a future direction, we plan to extend this study through industrial case studies to assess how these models perform in practical software development settings. This will help bridge the gap between benchmark-based evaluations and real-world deployments.


\section*{Acknowledgments}
This work has been partially supported by FAST, the Finnish Software Engineering Doctoral Research Network, funded by the Ministry of Education and Culture, Finland.

\section*{Declaration of AI Assistance}
During the preparation of this work, the author(s) used ChatGPT to refine grammar, improve sentence structure, and resolve formatting issues. After utilising this tool, the author(s) thoroughly reviewed and edited the content as needed, taking full responsibility for the final publication.

\printcredits

\bibliographystyle{elsarticle-num-names}
\bibliography{references}

\balance
\end{sloppypar}
\end{document}